\documentclass[preprint2,twocolumn,times,tighten]{aastex63}

\usepackage{graphicx,times}
\usepackage{subfigure}
\newcommand{\be}{\begin{equation}}
\usepackage{threeparttable}
\usepackage{booktabs}
\newcommand{\ee}{\end{equation}}
\newcommand{\bea}{\begin{eqnarray}}
\newcommand{\eea}{\end{eqnarray}}

\usepackage{enumerate}
\usepackage{amsmath}
\usepackage{cases}
\usepackage{longtable}
\usepackage{hyperref}
\usepackage{epstopdf}
\usepackage{amsmath,bm}
\usepackage{amssymb}
\usepackage{natbib}
\usepackage{morefloats}
\usepackage{multirow}
\usepackage{array}
\usepackage{verbatim}
\usepackage{lineno}

\shorttitle{Measuring magnetization with rotation measures and velocity centroids }
\shortauthors{Xu \& Hu}

\begin{document}

\title{Measuring magnetization with 
rotation measures and velocity centroids in supersonic MHD turbulence}

\email{sxu@ias.edu; yue.hu@wisc.edu}

\author[0000-0002-5771-2055]{Siyao Xu}
\affiliation{Institute for Advanced Study, 1 Einstein Drive, Princeton, NJ 08540, USA\footnote{Hubble Fellow}}
\author[0000-0002-8455-0805]{Yue Hu}
\affiliation{Department of Physics, University of Wisconsin-Madison, Madison, WI 53706, USA}
\affiliation{Department of Astronomy, University of Wisconsin-Madison, Madison, WI 53706, USA}

\begin{abstract}

The interstellar turbulence is magnetized and thus anisotropic. 
The anisotropy of turbulent magnetic fields and velocities is imprinted in the related observables, rotation measures (RMs) and velocity centroids (VCs). This anisotropy provides valuable information on both direction and strength of the magnetic field. 
However, its measurement is difficult especially in highly supersonic turbulence in cold interstellar phases due to the distortions by isotropic density fluctuations. By using 3D simulations of supersonic and sub-Alfv\'{e}nic magnetohydrodynamic (MHD) turbulence, we find that the problem can be alleviated when we selectively sample the volume-filling low-density regions in supersonic MHD turbulence. 
Our results show that 
in these low-density regions, 
the anisotropy of 
RM and VC fluctuations depends on the Alfv\'{e}nic Mach number as $\rm M_A^{-4/3}$. 
This anisotropy-$\rm M_A$ relation is
theoretically expected for sub-Alfv\'{e}nic MHD turbulence and 
confirmed 
by our synthetic observations of $^{12}$CO emission. 
It provides a new method for measuring the {plane-of-the-sky} magnetic fields in cold interstellar phases. 
\end{abstract}


\section{Introduction}

The interstellar medium (ISM) is both turbulent and magnetized, and turbulence and magnetic fields are dynamically coupled together. Understanding their properties is crucial for studying many multi-scale physical processes in the multi-phase ISM, 
including star formation,
cosmic ray propagation, 
and turbulent dynamo
\citep{ElmegreenScalo,2007ARA&A..45..565M}.

Despite their significance, measuring interstellar turbulence and magnetic fields is difficult. 
One of the main reasons for the difficulty is the involvement of densities in observables 
of turbulent velocities and magnetic fields. 
Some techniques aimed at disentangling contributions of turbulent velocities and densities, 
such as the 
Velocity Channel Analysis (VCA)
\citep{LP00},
the Velocity Coordinate Spectrum (VCS)
\citep{LP06},
have been developed and applied to extracting statistical properties of turbulent velocities in the ISM
from spectroscopic observations 
\citep{Laz09rev,Chep10}.

Turbulence in the presence of magnetic fields is anisotropic. 
The turbulent energy cascade mainly occurs in the direction perpendicular to the local magnetic field, 
resulting in larger velocity fluctuations in the perpendicular direction
\citep{GS95,LV99}.
\footnote{The importance of the reference frame of the local magnetic field for MHD turbulence was first discussed
in \citet{LV99}.}
With the decrease of turbulent velocity along the turbulent energy cascade, the anisotropy becomes more pronounced at smaller scales. The scale-dependent anisotropy becomes nearly scale independent when the measurement is performed in the global reference system with respect to the mean magnetic field, as demonstrated by both simulations 
\citep{CV00,MG01,CLV_incomp,Bere15} and observations of the solar wind (e.g., \citealt{Mat90,Luo10,Wi11}). Only the anisotropy in the global reference frame is accessible to observations of the ISM, which is the anisotropy averaged along the line-of-sight (LOS). While both turbulent velocities and magnetic fields  are anisotropic
\citep{LP12}, the density fluctuations in highly supersonic MHD turbulence  generated by shock compression are isotropic 
\citep{CL04,BLC05}. Consequently, retrieving  turbulence anisotropy from observables involving densities in highly supersonic turbulence in cold interstellar phases, e.g., molecular clouds, is very challenging. 

Based on the theoretical understanding on anisotropy of turbulent velocities, new techniques, such as the Velocity Gradients Technique 
(VGT, e.g., \citealt{Gonz17,Yu17,2018MNRAS.480.1333H}), 
and the Principal Component Analysis (e.g., \citealt{Hey08})
have been introduced for measuring interstellar turbulence and magnetic fields.  Statistical studies of velocity centroids (VCs, 
e.g., \citealt{LE03,EL05,BurL14,Kan17}) show that VCs can be used to probe the  turbulence anisotropy and the {plane-of-the-sky (POS)} magnetic fields. However, VC fluctuations in highly supersonic MHD turbulence are dominated by density fluctuations and are unable to fully reveal the anisotropy of turbulent velocities \citep{Hu20}. 
The specific dependence of VC anisotorpy on magnetic field strength is unclear. 

Similarly, the commonly used observables for tracing magnetic fields, e.g., Faraday rotation measures (RMs), are also subject to the distortions by densities. Statistical analysis of interstellar RM fluctuations \citep{MS96} suggests that they are dominated by density fluctuations \citep{XZ16}, which exhibit a shallow density spectrum due to the small-scale density enhancements arising from the supersonic turbulence in cold interstellar phases. This finding on a shallow density spectrum is also confirmed by the temporal broadening measurements \citep{XuZ17} and dispersion measures \citep{Xup20} of pulsars. This finding is also consistent with the shallow density spectra measured with gas and dust tracers in the cold neutral medium and molecular clouds  \citep{HF12}, showing the coupling between ionized and neutral components of gas in the partially ionized ISM 
\citep{XLY14,Xuc16}.

\citet{Hua20} investigated the measurement on orientation and strength of magnetic fields from anisotropic turbulent velocities, which can be directly obtained by using point tracers of turbulence, including 
dense cores \citep{Qi18} and young stars \citep{Ha20}. The dependence of turbulence anisotropy on the Alfv\'{e}nic Mach number $\rm M_A$, which is related to the magnetic field strength, was analytically derived and numerically tested by \citet{Hua20}. It leads to a new method for measuring magnetic fields in sub-Alfv\'{e}nic turbulence ($\rm M_A<1$) in the ISM. In this work, we consider more commonly used observables for tracing turbulent magnetic fields and velocities, i.e., RMs and VCs, which are subject to both projection effect and distortions by density fluctuations. By exploring a new analysis method to mitigate the effect of densities and maximize the information on turbulence anisotropy, we will reexamine their applicability to supersonic MHD turbulence in  retrieving turbulence anisotropy and measuring magnetization. We will also test our results on VCs by using synthetic observations including optical depth effects.

In Section~\ref{sec:2}, we present the theoretical formulations for structure functions (SFs) of turbulent fluctuations, which are a basic statistical tool for analyzing turbulence and quantifying turbulence anisotropy. In Section~\ref{sec:3}, we numerically study the anisotropy of SFs of RMs and VCs by using a set of MHD simulations of supersonic and sub-Alfv\'{e}nic turbulence and produce synthetic observations for testing our results. The {discussion and} conclusions can be found in {Section~\ref{sec:4} and Section~\ref{sec:5}}. 

\section{SF analysis in MHD turbulence}
\label{sec:2}
\subsection{Anisotropy of MHD turbulence}

Magnetized turbulence is anisotropic, and the anisotropy 
becomes more pronounced toward smaller scales with the turbulent energy cascade. Here we focus on the strong turbulence regime where the critical balance between the turbulent eddy-turnover time and 
the Alfv\'{e}n wave period is established  \citep{GS95}.

Following the turbulent energy cascade, the Kolmogorov scaling for hydrodynamic turbulence still applies to MHD turbulence in the direction perpendicular to the local magnetic field, as numerically demonstrated with high-resolution MHD turbulence simulations in, e.g.,
\citet{Bere14}. Therefore, the turbulent velocity at the length scale $l_\perp$ perpendicular to the local magnetic field is given by 
\begin{equation}\label{eq: kol}
   v (l_\perp)= V_\text{st} \Big(\frac{l_\perp}{L_\text{st}}\Big)^\frac{1}{3},
\end{equation}
where $V_\text{st}$ is the turbulent velocity at $L_\text{st}$. For sub-Alfv\'{e}nic turbulence with the turbulent energy smaller than the magnetic energy, the Alfv\'{e}nic Mach number $M_A = V_L / V_A$ is less than unity, where $V_L$ is the injected turbulent velocity at the injection scale $L_i$, $V_A = B / \sqrt{4 \pi\rho}$ is the Alfv\'{e}n speed, $B$ is the magnetic field strength, and $\rho$ is the density. 
There is 
\citep{Lazarian06}
\begin{equation}\label{eq: sub}
    V_\text{st} = V_L M_A, ~ L_\text{st} = L_i M_A^2,
\end{equation}
and strong MHD turbulence exists on scales smaller than $L_\text{st}$. 
The scale-dependent turbulence anisotropy is described by 
\begin{equation}\label{eq: locani}
     l_\|= \frac{V_A}{V_\text{st}} L_\text{st}^\frac{1}{3} l_\perp^\frac{2}{3}, 
\end{equation}
which can be derived by combining Eq. \eqref{eq: kol} with the critical balance relation $l_\perp / v_l  = l_\| / V_A$, 
where $l_\|$ is the parallel length scale measured along the local magnetic field. It shows that the turbulence becomes more anisotropic at smaller scales. 

Given the relation in Eq. \eqref{eq: locani}, one can express $v$ in Eq. \eqref{eq: kol} in terms of $l_\|$, 
\begin{equation}
    v (l_\|) = V_\text{st} \Big(\frac{V_\text{st}}{V_A}\Big)^\frac{1}{2} \Big(\frac{l_\|}{L_\text{st}}\Big)^\frac{1}{2}.
\end{equation}
Comparing the above expression with Eq. \eqref{eq: kol}, when $v(l_\perp)$ and $v(l_\|)$ are measured at the same length, i.e., $l_\perp = l_\|$, we define the anisotropy degree (AD) in the system of reference aligned with the local magnetic field as  
\begin{equation}
   \text{AD}_\text{loc} = \frac{v (l_\perp)^2}{ v(l_\|)^2} =  \frac{V_A}{V_\text{st}} \Big(\frac{l_\perp}{L_\text{st}}\Big)^{-\frac{1}{3}}.
\end{equation}
Inserting Eq. \eqref{eq: sub} into it yields 
\begin{equation}
    \text{AD}_\text{loc} = M_A^{-\frac{4}{3}} \Big(\frac{l_\perp}{L}\Big)^{-\frac{1}{3}},   ~~    (M_A < 1),
\end{equation}
which increases with decreasing $l_\perp$.

The above scale-dependent anisotropy can only be measured in the local reference frame aligned with the locally averaged magnetic field direction \citep{LV99,CV00,CLV_lecnotes}. Observations in the ISM are subjected to the projection effect, and only the global reference frame of the mean magnetic field averaged along the LOS is accessible.
The measured anisotropy is dominated by that of the largest turbulent eddy in the sampled volume, as the local anisotropies of smaller eddies are averaged out. The anisotropic scaling of the largest eddy at $L_\text{st}$ is given by (Eq. \eqref{eq: locani})
\begin{equation}
     L_{\text{st},\|} \approx \frac{V_A}{V_\text{st}} L_{\text{st},\perp}, 
\end{equation}
where we assume $L_\text{st} \approx L_{\text{st},\perp}$ for anisotropic turbulence. The scale-independent anisotropy in the 
global frame is 
\begin{equation}
     \frac{l_\|}{l_\perp} = \frac{L_{\text{st},\|}}{L_{\text{st},\perp}} \approx \frac{V_A}{V_\text{st}}. 
\end{equation}
Then Eq. \eqref{eq: kol} can be expressed in terms of $l_\|$, 
\begin{equation}\label{eq: vpaglo}
     v (l_\|) \approx V_\text{st} \Big(\frac{ l_\| }{L_{\text{st},\|}}\Big)^\frac{1}{3}.
\end{equation}
The AD in the global frame is the ratio 
(Eqs. \eqref{eq: kol} and \eqref{eq: vpaglo})
\begin{equation}
   \text{AD}_\text{glo} =  \frac{v(l_\perp)^2}{ v(l_\|)^2} \approx \Big(\frac{L_{\text{st},\|}}{L_\text{st}}\Big)^\frac{2}{3}
    \approx \Big(\frac{V_A}{V_\text{st}}\Big)^\frac{2}{3}
\end{equation}
measured at  $l_\perp = l_\|$, where $l_\perp$ and $l_\|$ are the perpendicular and parallel length scales measured with respect to the direction of the mean magnetic field. By using Eq. \eqref{eq: sub}, we further get 
\begin{equation}\label{eq: 3danig}
   \text{AD}_\text{glo} \approx M_A^{-\frac{4}{3}}, ~~    (M_A < 1).
\end{equation}
Both AD$_\text{loc}$ and AD$_\text{glo}$ have the same dependence on $\rm M_A$ and have larger values for more strongly magnetized turbulence with a smaller $\rm M_A$. But in contrast to AD$_\text{loc}$, AD$_\text{glo}$ is scale-independent. 

The above anisotropic scaling for the turbulent velocities in 
incompressible MHD turbulence is also applicable to compressible MHD turbulence, when the turbulent motions are governed by Alfv\'{e}n modes \citep{CL02_PRL,CL03}. 

\subsection{Anisotropy of SFs of turbulent fluctuations}
For the statistical analysis of turbulent fluctuations carried out below, we consider the statistics in the global frame of reference as this is the only available statistics for observations of the ISM. 
\subsubsection{3D SF}
For the statistical measurement of the fluctuations induced by turbulence, when the three-dimensional (3D) positions are available, we have the 3D correlation function (CF), 
\begin{equation}
\begin{aligned}
    \xi_v(R,\Delta z) &= \langle v(\bm{X_1},z_1) v(\bm{X_2},z_2)\rangle \\
    &= \langle v^2 \rangle \frac{L_\text{st}^m}{L_\text{st}^m  +  (R^2 + \Delta z^2)^\frac{m}{2}}  ,
\end{aligned}
\end{equation}
and 3D SF 
\begin{equation}\label{eq: trdsfsd}
\begin{aligned}
      d_v(R, \Delta z) &= \langle [v(\bm{X_1},z_1) - v(\bm{X_2},z_2) ]^2 \rangle \\
      & =2\langle v^2 \rangle \frac{(R^2 + \Delta z^2)^\frac{m}{2}}{L_\text{st}^m  +  (R^2 + \Delta z^2)^\frac{m}{2}}  .
\end{aligned}
\end{equation}
We adopt the power-law models for the CF and SF \citep{LP16}, with the power-law index $m = 2/3$ corresponding to the Kolmogorov scaling of turbulence. They are applied to turbulent velocities in the above expressions as an example. Here $\bm{X}$ and $z$ are the projected position vector on the plane of sky and the distance along the LOS, 
$R = |\bm{X_1} - \bm{X_2}|$, $\Delta z = z_1 - z_2$, and $\langle...\rangle$ denotes the average over all pairs of points with the same 3D separation. 

The AD of the 3D SF of turbulent velocities is given by Eq. \eqref{eq: 3danig},
\begin{equation}
\label{eq: tdsfanima}
     \frac{d_v(r_\perp)}{d_v(r_\|)} = \text{AD}_\text{glo} \approx M_A^{-\frac{4}{3}},
\end{equation}
where we use $r_\perp$ and $r_\|$ to represent the separations measured in the directions perpendicular and parallel to the mean magnetic field. The above relation has been numerically tested by 
\citet{Hua20}. Their finding suggests that by measuring the 3D SF of turbulent velocities along different directions, one can determine both the orientation of the mean magnetic field and the value of $\rm M_A$. The latter leads to the evaluation of $B$ when $\rho$ is known. 
This method can be applied to young stars with full 6D phase-space coordinates provided by Gaia, which have been recently found as a new probe of turbulence in their parent molecular clouds \citep{Ha20}.  

\begin{figure*}
\centering
\subfigure[$L \ll L_\text{st}$]{
   \includegraphics[width=8cm]{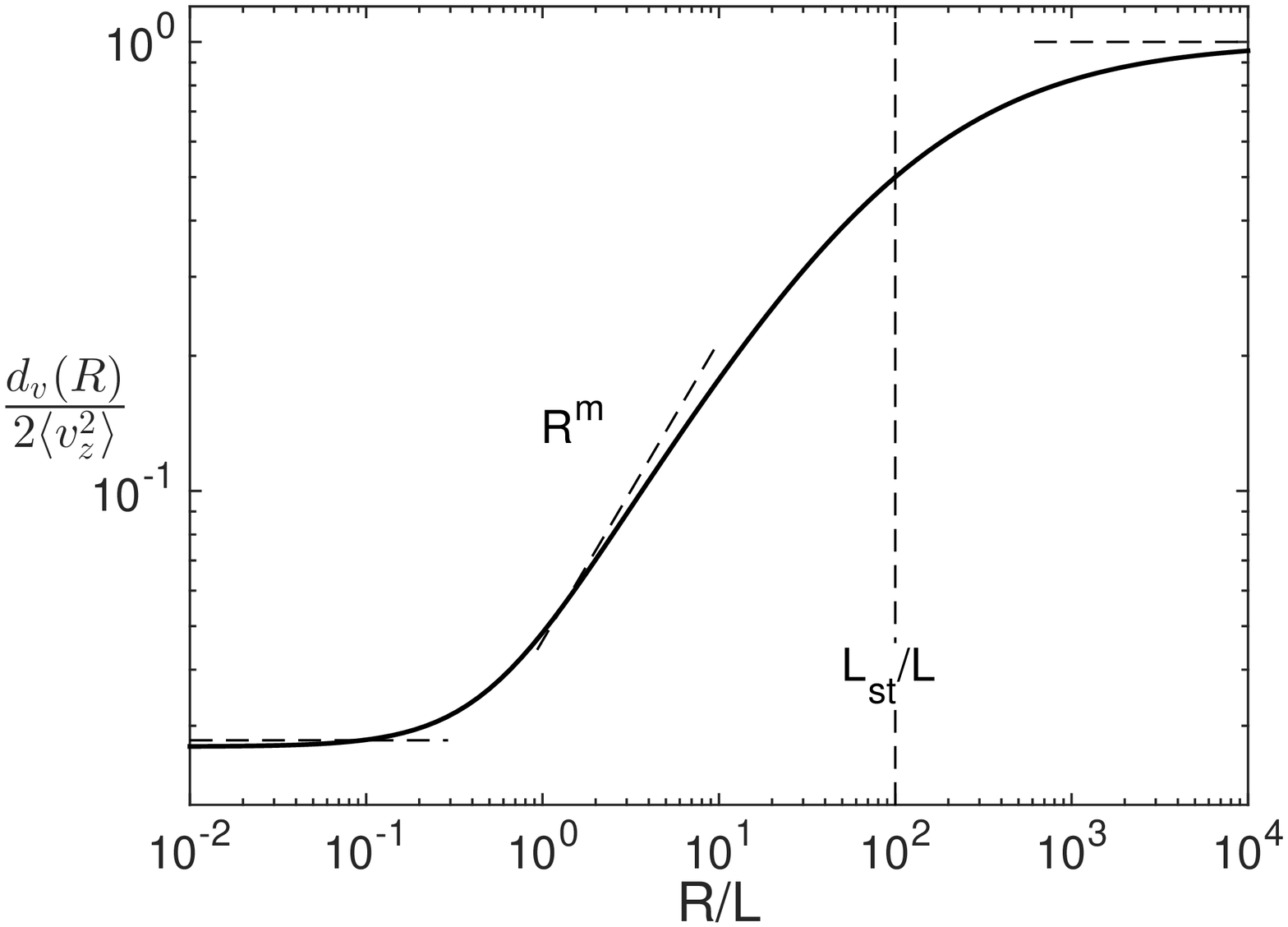}\label{fig: thin}}
\subfigure[$L \gg L_\text{st}$]{
   \includegraphics[width=8cm]{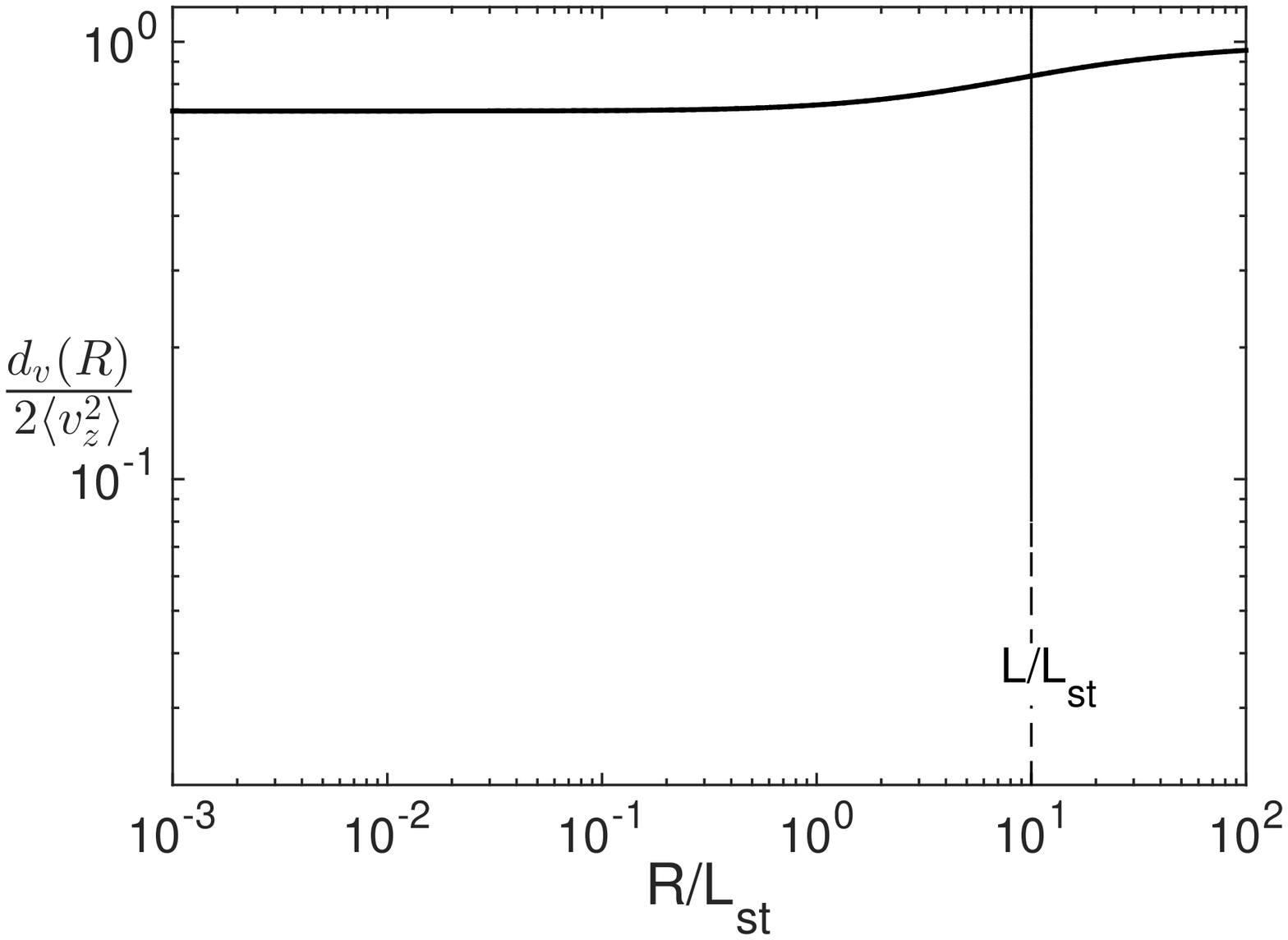}\label{fig: thick}}
\caption{Projected 3D SF $d_v(R)$ (normalized by $2 \langle v_z^2 \rangle$) of turbulent velocities 
sampled by point sources in the (a) thin and (b) thick turbulent volumes. }
\label{fig: illu}
\end{figure*}

For some other point tracers of turbulence, such as dense cores in molecular clouds \citep{Qi12}, only the LOS component of turbulent velocity $v_z$ and the 2D position on the plane of sky are attainable.
In this case, the 3D SF is averaged over all pairs with the same $R$ but different $\Delta z$, 
\begin{equation}
\begin{aligned}
    d_v(R) &= \frac{\int_0^L d_v(R, \Delta z) d \Delta z }{\int_0^L    d \Delta z} \\
            &= \frac{2\langle v_z^2 \rangle}{L}  \int_0^L   \frac{(R^2 + \Delta z^2)^\frac{m}{2}}{L_\text{st}^m  +  (R^2 + \Delta z^2)^\frac{m}{2}}   d \Delta z ,
\end{aligned}
\end{equation}
where we assume that the point sources are distributed uniformly along the LOS 
from $z=0$ to $z=L$, and $L$ is the thickness of the sampled volume. 
As illustrated in Fig.~\ref{fig: thin},
when the turbulent volume is thin with $ L \ll L_\text{st} $, there is
\begin{equation}
\begin{aligned}
  &  d_v(R) \approx 2\langle v_z^2 \rangle   \frac{1}{m+1}   \Big(\frac{L}{L_\text{st}}\Big)^m, ~~R < L, \\
  &  d_v(R) \approx 2\langle v_z^2 \rangle      \Big(\frac{R}{L_\text{st}}\Big)^m, ~~~~~ L< R < L_\text{st}, \\
  &  d_v(R) \approx 2\langle v_z^2 \rangle, ~~~~~~~~~~~~~~~~~~~~~~~~~~ R> L_\text{st}. 
\end{aligned}
\end{equation}
Both the scaling and anisotropy properties of turbulence can be retrieved from $d_v(R)$ within the range $ L<R <L_\text{st}$ (see \citealt{Hua20}). When the LOS is perpendicular to the mean magnetic field, we have 
\begin{equation}
    \frac{d_v(R_\perp)}{d_v(R_\|)} = \text{AD}_\text{glob }\approx M_A^{-\frac{4}{3}},
\end{equation}
where $R_\perp$ and $R_\|$ are measured with respect to the direction of the mean magnetic field.

However, when the turbulent volume has the LOS thickness much larger than the correlation length $L_\text{st}$, 
there is always 
\begin{equation}
    d_v(R) \approx 2\langle v_z^2 \rangle .
\end{equation}
The properties of turbulence cannot be retrieved (see Fig.~\ref{fig: thick} and also \citealt{Hua20}). This result can be potentially used for constraining the thickness of molecular clouds \citep{Qian15}.

\subsubsection{2D SF}
\label{sssec:2dsf}
When using spatially continuous gas tracers of turbulence, 
we usually need to deal with projected quantities, i.e., integrated quantities along the LOS, and their SFs. In the case of turbulent velocities, we have the 2D SF of projected velocities as 
\citep{LP16}
\begin{equation}\label{eq: drpjt}
\begin{aligned}
   D_v(R) &= \langle [v_z(\bm{X_1}) - v_z(\bm{X_2}) ]^2 \rangle\\
          &= \Big\langle \Big[ \int_0^L dz v_z(\bm{X_1},z)- \int_0^L dz v_z(\bm{X_2},z)  \Big] ^2 \Big\rangle\\
          & = 4 \int_0^L d \Delta z (L - \Delta z) [\xi_v(0, \Delta z) - \xi_v(R, \Delta z)] \\
          & = 4  \langle v_z^2 \rangle \int_0^L d\Delta z (L-\Delta z)  \\
          & ~~~~~~ \Bigg[    \frac{L_\text{st}^m}{L_\text{st}^m  +  \Delta z^m} -    \frac{L_\text{st}^m}{L_\text{st}^m  +  (R^2 + \Delta z^2)^\frac{m}{2}}  \Bigg] ,
\end{aligned}
\end{equation}
where the LOS is in the $z$ direction, and $L$ is the thickness of the turbulent gas cloud. For a thick cloud with $L$ larger than $L_\text{st}$ and for Kolmogorov scaling of turbulent velocities, 
$D_v(R)$ can be approximated by 
\begin{equation}
\label{eq: app1thick}
   D_v(R)\approx 4  \langle v_z^2 \rangle \int_0^R d\Delta z L \frac{R^m}{L_\text{st}^m}  
    = 4  \langle v_z^2 \rangle L_\text{st}^{-m} L  R^{m+1} 
\end{equation}
within the range $R < L_\text{st}$. Compared with 3D SF in Eq. \eqref{eq: trdsfsd}, $D_v(R)$ has a steeper power-law dependence on $R$ because of the projection effect, as shown in observations \citep{LP00,Elm01,Pad01}. 

The detailed analysis on the anisotropy of $D_v(R)$ of different MHD modes and its relation to the magnetization and the angle between LOS and mean magnetic field was carried out by \citet{Kan17}. Here as a rough estimate, we rewrite $D_v$ in terms of $R_\perp$ and $R_\|$, which are the 2D projected separations measured perpendicular and parallel to the mean magnetic field, and have 
\begin{equation}
   D_v(R_\perp)     \approx 4  \langle v_z^2 \rangle \int_0^{R_\perp} d\Delta z L  
                   \frac{R_\perp^m}{L_\text{st}^m  }  
                = 4  \langle v_z^2 \rangle L_\text{st}^{-m} L  R_\perp^{m+1} ,
\end{equation}
and 
\begin{equation}
   D_v(R_\|)     \approx 4  \langle v_z^2 \rangle \int_0^{R_\|} d\Delta z L  
                   \frac{R_\|^m}{L_{\text{st},\|}^m  }  
                = 4  \langle v_z^2 \rangle L_{\text{st},\|}^{-m} L  R_\|^{m+1} , 
\end{equation}
where we assume that the mean magnetic field is perpendicular to the LOS for simplicity. Then the AD measured with 2D SFs at $R_\perp = R_\|$ can be estimated as 
\begin{equation}
     \frac{D_v(R_\perp) }{D_v(R_\|) } \approx \Big(\frac{L_{\text{st},\|}}{L_\text{st}}\Big)^m = M_A^{-2m},
\end{equation}
which has the same dependence on $\rm M_A$ as AD$_\text{glo}$ measured with 3D SFs (Eq. \eqref{eq: tdsfanima})
for Kolmogorov scaling of turbulent velocities.

The above analysis on anisotropy of SFs is exemplified by turbulent velocities as 
they directly reflect the dynamics of MHD turbulence and are the best indicator of turbulence anisotropy. 
The anisotropy of turbulent magnetic fields and its application to analyzing the 
statistics of synchrotron fluctuations 
have been comprehensively studied by 
\citet{LP12}. 
Magnetic fields are mainly perturbed by Alfv\'{e}nic turbulent motions 
even in highly supersonic turbulence and thus follow similar anisotropic properties as turbulent velocities, 
as shown in numerical simulations with solenoidal driving 
\citep{BLC05}. 
Density fluctuations in subsonic and mildly supersonic MHD turbulence
are passively mixed by Alfv\'{e}nic motions and have the same anisotropy as turbulent velocities
\citep{LG01,CL03, XJL19}. 
While in highly supersonic turbulence, high density contrast is generated due to shock compression, 
and density fluctuations become isotropic 
\citep{CL04, BLC05, Hu20,2020ApJ...897..123H}.

Magnetic fields and velocities are usually coupled with densities in observational measurements, 
and the extraction of them is very challenging especially in highly supersonic turbulence. 
The commonly used observable for studying magnetic fields is RM, which is defined as 
\begin{equation}
     \text{RM} = \frac{e^3}{2\pi m_e^2 c^4} \int_0^Ldz n_e B_z ,
\end{equation}
where $e$ and $m_e$ are the charge and mass of an electron, $c$ is the light speed, 
$n_e$ is the number density of electrons, and $B_z$ is the LOS component of magnetic field.
Turbulent magnetic fields and densities both contribute to the fluctuations of RMs, but the latter
can play a dominant role due to the large density variation arising from supersonic turbulence. 
The resulting SF of RMs in the ISM measured by, e.g., 
\citet{MS96}, 
can be explained by a shallow density spectrum expected for supersonic turbulence in cold interstellar phases
\citep{XZ16}.

The VC used for measuring turbulent velocities can be defined as 
\begin{equation}
     \text{VC} = \epsilon \int_0^L dz \rho v_z 
\end{equation}
in position-position-position space for optically thin emission lines 
\citep{Kan17},
where $\epsilon$ is the emissivity coefficient, and 
$\rho$ is the real space density. 
Similar to RMs, VCs do not exhibit 
the scaling properties of the underlying turbulent velocities in highly supersonic turbulence
due to the density distortions
\citep{EL05,Esq07}.

The SFs of RMs and VCs are 
\begin{equation}
    D_\text{RM} (R) = \langle [\text{RM}(\bm{X_1}) - \text{RM}(\bm{X_2})]^2 \rangle,
\end{equation}
and 
\begin{equation}
    D_\text{VC} (R) = \langle [\text{VC}(\bm{X_1}) - \text{VC}(\bm{X_2})]^2 \rangle,
\end{equation}
respectively. 
We will next carry out numerical simulations and investigate 
whether they can be used to recover the turbulence anisotropy in 
supersonic and sub-Alfv\'{e}nic MHD turbulence.

\section{Anisotropy of SFs of RMs and VCs}
\label{sec:3}

\subsection{Simulations of supersonic and sub-Alfv\'{e}nic MHD turbulence}

\label{sec:data}

Our scale-free 3D MHD simulations are generated using the ZEUS-MP/3D code \citep{2006ApJS..165..188H}, which solves the ideal MHD equations with zero-divergence condition $\nabla \cdot\Vec{B}=0$ and an isothermal equation of state in a periodic box. Here $\Vec{B}$ is the total magnetic field and the box is regularly staggered to 792$^3$ grid cells. We consider single fluid and operator-split MHD conditions in the Eulerian frame. The turbulence is solenoidally injected in Fourier space at {wavenumber $k_i\approx$ 2, and 
the dissipation occurs near $k_d\approx100$ 
(see \citealt{Hu20})}. 
The simulations are initialized with a uniform density and a uniform magnetic field $\Vec{B_0}$.  
The magnetic field consists of $\Vec{B_0}$ and a fluctuating component $\Vec{b}$, so $\Vec{B}=\Vec{B_0}+\Vec{b}$.
{Initially we set the magnetic field along the $x$ and $z$ axes to be zeros, and the uniform magnetic field is along the $y$-axis. We choose $z$-axis to be the LOS so that the mean magnetic field is perpendicular to the LOS.}

We consider supersonic and sub-Alfv\'{e}nic MHD turbulence for studying the anisotropy of supersonic turbulence in cold molecular clouds (MCs) with the sonic Mach number $M_S = V_L/ c_s \approx 5-20$ \citep{Zuc74,Lars81}, where $c_s$ is the sound speed. The values of $\rm M_S$ and $\rm M_A$ of our numerical models are listed in Table \ref{tab:num}. For comparison, we also include one model (M0) with subsonic and sub-Alfv\'{e}nic turbulence.

\begin{table*}[!htbp]
\renewcommand\arraystretch{1.5}
\centering
\begin{threeparttable}
\caption{
$\rm M_S$ and $\rm M_A$ values of our MHD simulations. {$B_0$ is the initial magnetic field strength corresponding to mean mass density $\langle\rho\rangle=300$ $g$ $cm^{-3}$ and isothermal sound speed $c_s$ = 187 $m$ $s^{-1}$.}
}
\label{tab:num} 
\begin{tabular}{c|c|c|c|c|c|c|c|c|c|c}
\toprule
Model & M0 & M1 & M2 & M3 & M4 & M5 & M6 & M7 & M8 & M9 \\
\hline
$\rm M_S$ & $0.62$ & $7.31$ & $6.10$ & $6.47$ & $6.14$ & $6.03$ & $10.81$ & $10.53$ & $10.61$ & $10.67$ \\\hline
$\rm M_A$ & $0.56$ & $0.22$ & $0.42$ & $0.61$ & $0.82$ & $1.01$ & $0.26$ & $0.51$ & $0.68$ & $0.95$ \\\hline
$B_0$ [$\mu G$] & 2.5 & 72 & 32 & 22 & 16 & 13 & 90 & 45 & 35 & 25 \\
\bottomrule
\end{tabular}
\end{threeparttable}
\end{table*}

Simulations by 
\citet{Pad16} 
suggest that the supernova diving is able
to sustain the turbulence observed in MCs, without invoking other sources of turbulence. 
The supernova energy injection brings both solenoidal and compressive modes, but 
the compressive-to-solenoidal ratio decreases due to the conversion from compressive to solenoidal motions along the 
energy cascade, leading to the dominance of solenoidal motions at MC scales. 
This finding is consistent with the observations showing the Kolmogorov scaling of turbulent velocities 
\citep{Qi18,Xu20}
and the dominance of solenoidal motions
\citep{Ork17}
at MC scales. 
Motivated by these numerical and observational results, we adopt 
the solenoidal driving for studying turbulence in MCs. 

For our measurements on anisotropic SFs, we select a 
sample of points for the 3D SFs and lines of sight for the 2D SFs from our data cube, 
and the separations are binned evenly with the bin width approximately equal to two grid cells
on scales smaller than $L_i$ ($\approx 792/k_i=396$ grid units). 
In the 2D case, we use a sample size of $10^4$. 
For both the SFs measured perpendicular and parallel to the mean magnetic field, 
the number of pairs in each separation bin is on the order of $10^3 - 10^4$, 
which is sufficiently large for the SF measurement to be statistically stable 
(see \citealt{Hua20}). 
{The uncertainties in the measured anisotropy of SFs related to the sample size will be presented 
in Sections \ref{sssec: lowd} and \ref{ssec:synobs}.}
For the measurement on 3D anisotropic SFs, as it requires a larger sample than the 2D case, 
we use a sample size of $10^6$ to have the number of pairs in each bin larger than $10^3$. 
To investigate the effect of density on the anisotropy of RM and VC SFs, 
we first randomly sample the turbulence volume in Section \ref{sssec: com}, 
and then selectively sample the low-density regions in Section \ref{sssec: lowd}.

\subsection{Numerical results}

\subsubsection{Comparisons between supersonic and subsonic MHD turbulence}
\label{sssec: com}

\begin{figure*}[htbp]
\centering   
\subfigure[Supersonic (M7)]{
   \includegraphics[width=8.5cm]{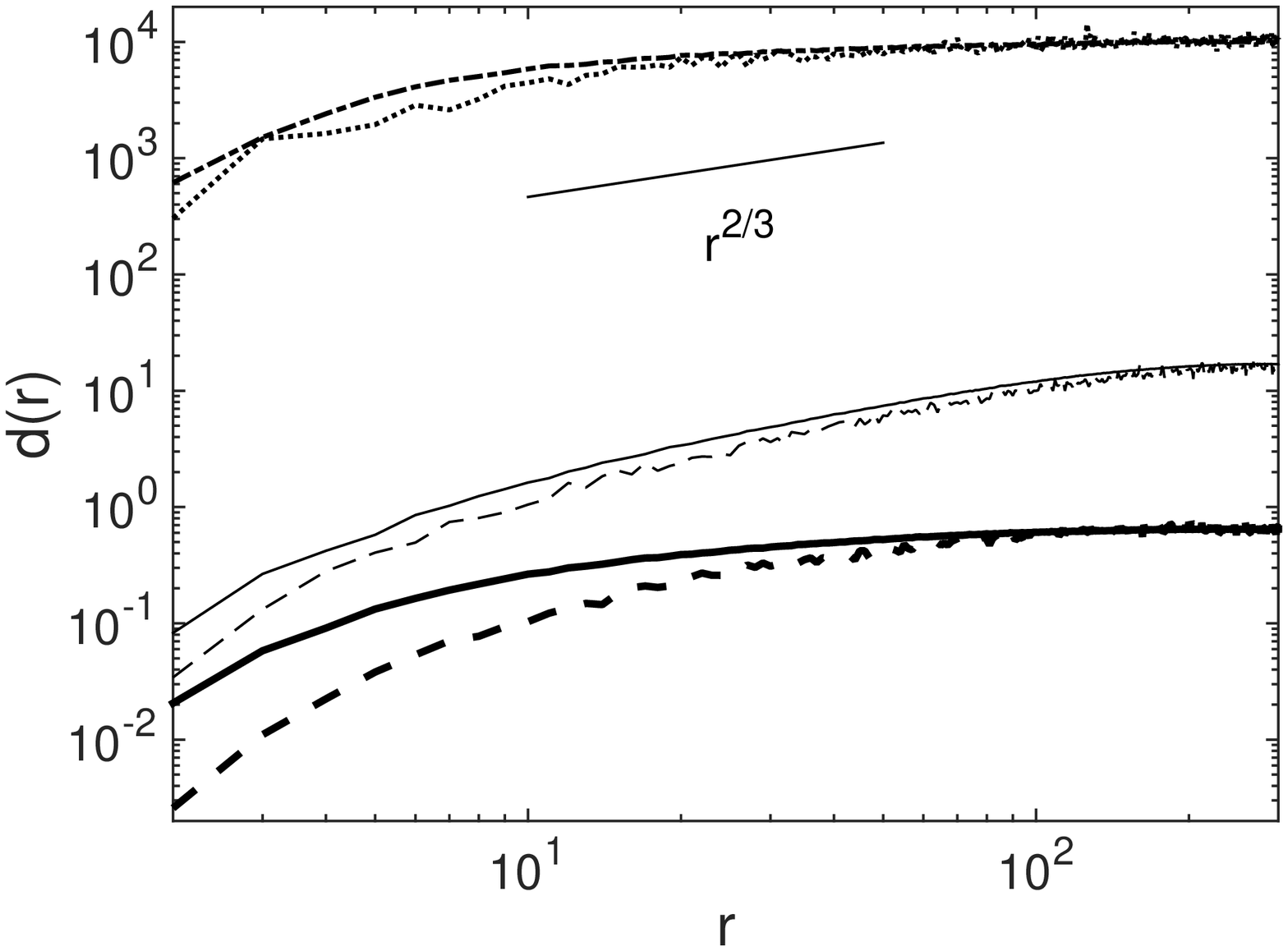}\label{fig: sup043d}}
\subfigure[Subsonic (M0)]{
   \includegraphics[width=8.5cm]{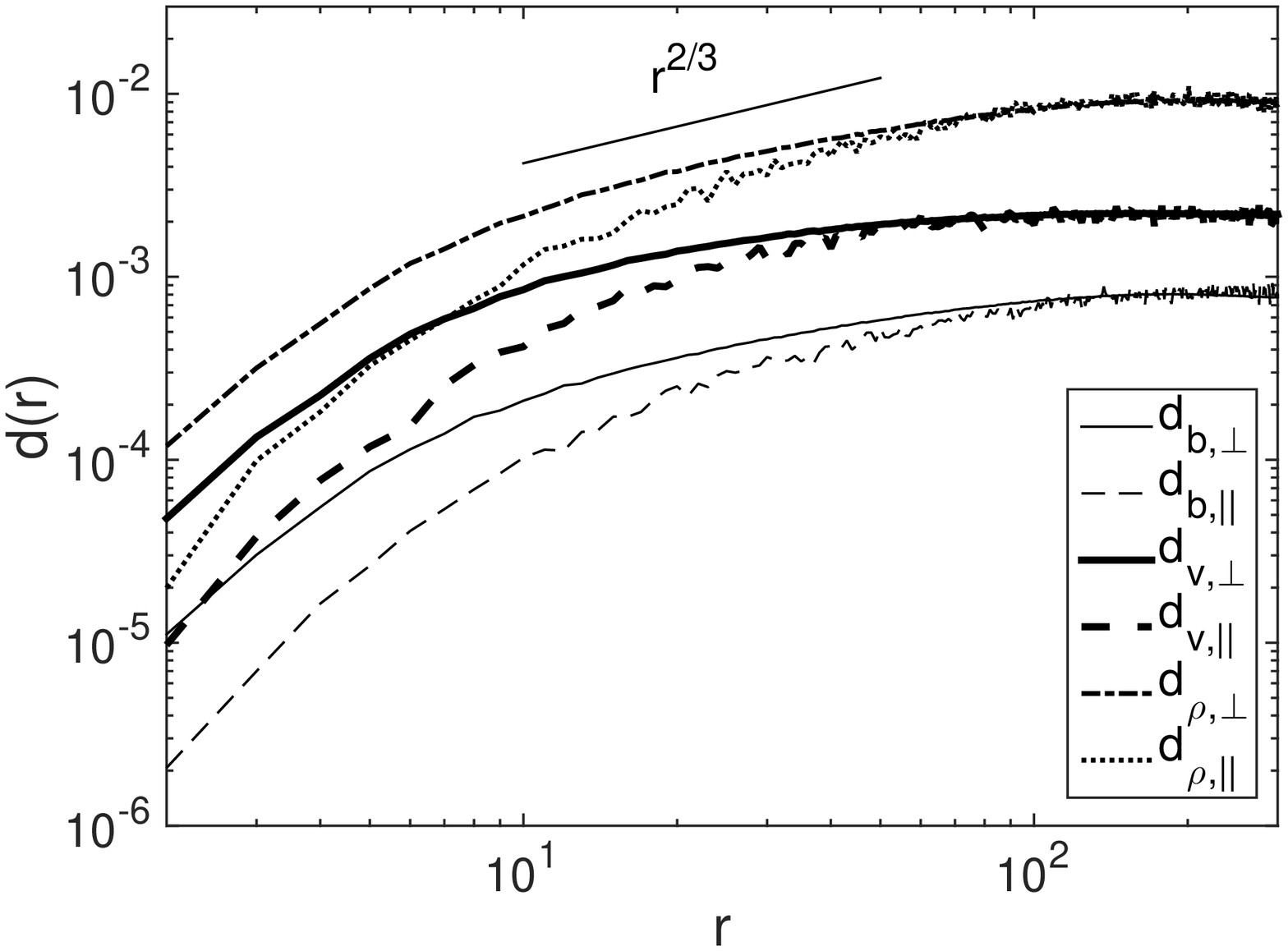}\label{fig: sub3d}}
\caption{(a) 3D SFs of turbulent magnetic fields, velocities, and densities measured for run M7. 
(b) Same as (a) but for M0. The same linestyles are used for both (a) and (b).
The Kolmogorov scaling is indicated by the short solid line. }
\label{fig: 3d}
\end{figure*}

Fig.~\ref{fig: 3d} presents the 3D SFs of turbulent magnetic fields $d_b$, velocities $d_v$, and densities $d_\rho$ in supersonic and subsonic MHD turbulence with similar $\rm M_A$ values. $d_{\perp}$ and $d_{\|}$ represent $d(r_\perp)$ and $d(r_\|)$ for simplicity. All units used in our measurements here and below are numerical units. 
{From the 3D SFs,
we see that the numerical dissipation effect sets in at $792/k_d \approx 8$
grid units, below which the SF steepens due to dissipation.} 
The difference between the SFs measured in the directions perpendicular and parallel to the mean magnetic field can be clearly seen in all cases for the subsonic MHD turbulence. The fake dependence of the anisotropy on length scales appears due to the isotropic driving and insufficient inertial range (see \citealt{CV00,Hua20}).
By contrast, we see in Fig.~\ref{fig: sup043d} that the SF of density fluctuations in highly supersonic turbulence is nearly isotropic as discussed in Section \ref{sssec:2dsf}. The different density structures in the supersonic and subsonic MHD turbulence are illustrated in Fig.~\ref{fig: illu}. Small-scale density enhancements with large $\rho / \langle \rho \rangle$ appear in the supersonic MHD turbulence, where $\langle \rho \rangle$ is the mean density,
in comparison with the more uniform density distribution in the subsonic case. 

\begin{figure*}[htbp]
\centering   
\subfigure[Supersonic (M7)]{
   \includegraphics[width=8cm]{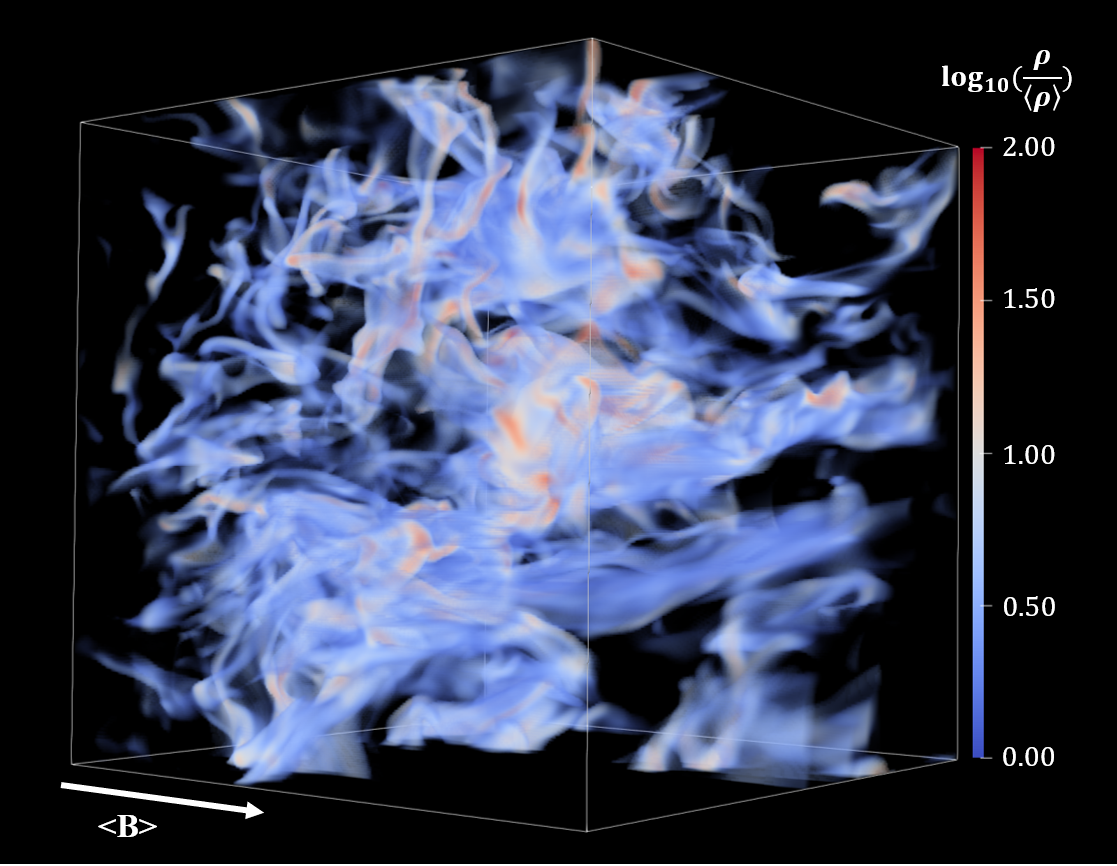}\label{fig: supden}}
\subfigure[Subsonic (M0)]{
   \includegraphics[width=8cm]{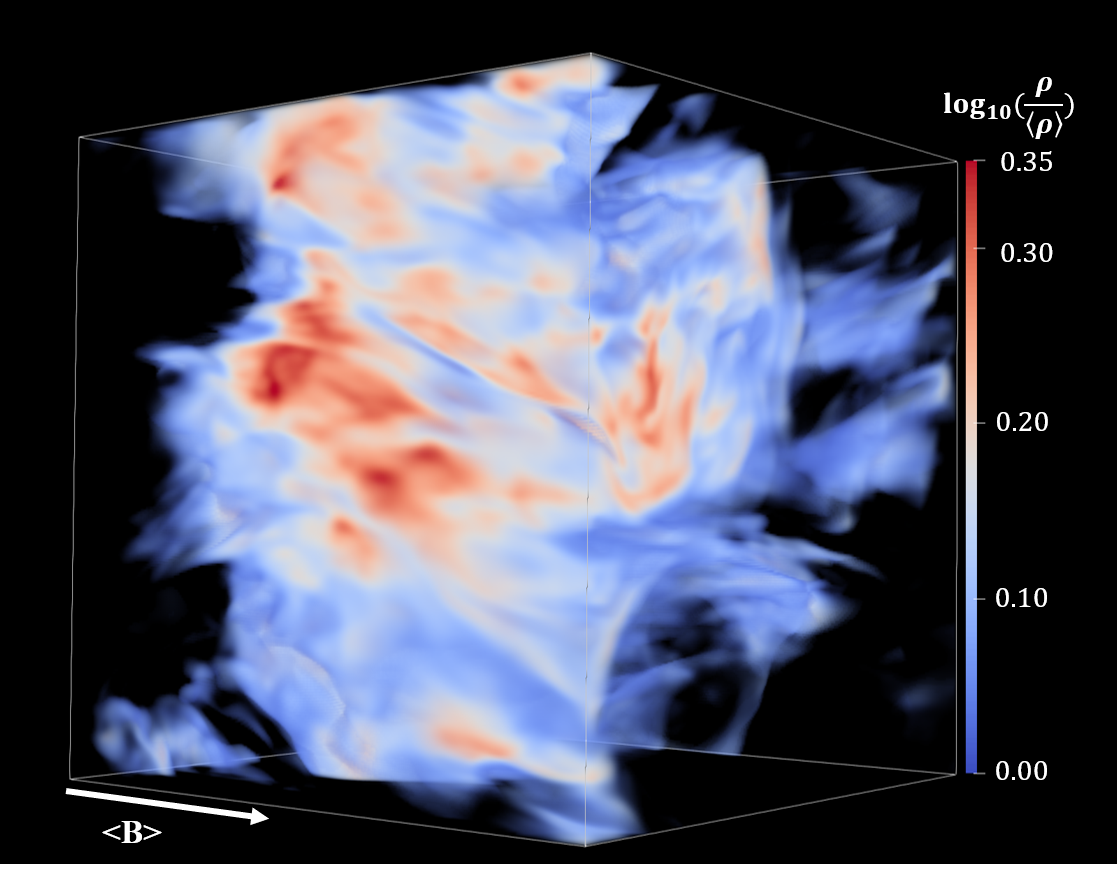}\label{fig: subden}}
\caption{Illustrations for the logarithmic density distribution 
(normalized by the mean density) in the supersonic MHD turbulence (run M7) and subsonic MHD turbulence (run M0). 
The direction of the mean magnetic field $\langle B \rangle$ is indicated by the arrow. 
Note that different color scales are used in (a) and (b).}
\label{fig: illu}
\end{figure*}

\begin{figure*}[htbp]
\centering   
\subfigure[]{
   \includegraphics[width=8.5cm]{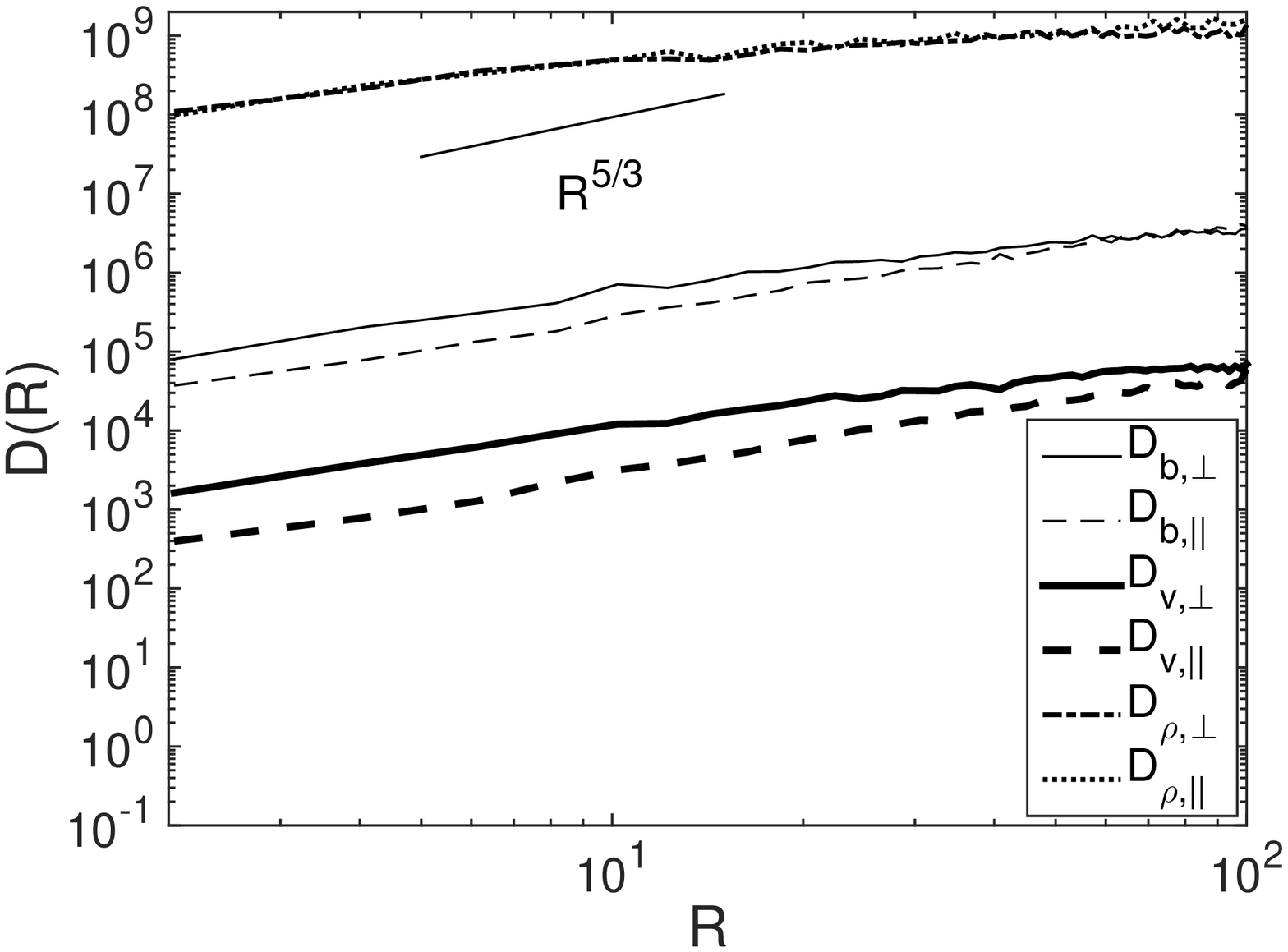}\label{fig: colsup4}}
\subfigure[]{
   \includegraphics[width=8.5cm]{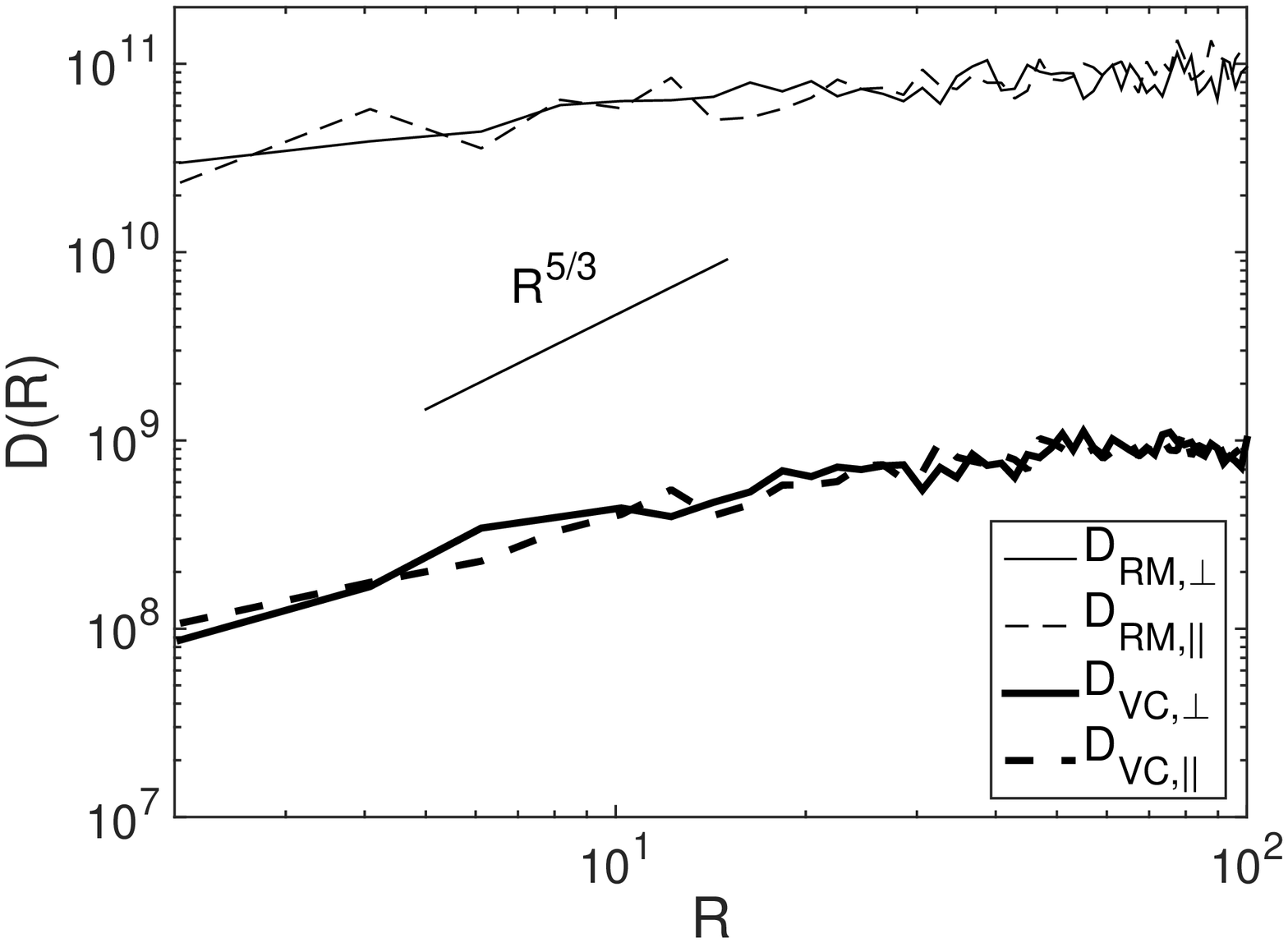}\label{fig: unsup4}}
\subfigure[]{
   \includegraphics[width=8.5cm]{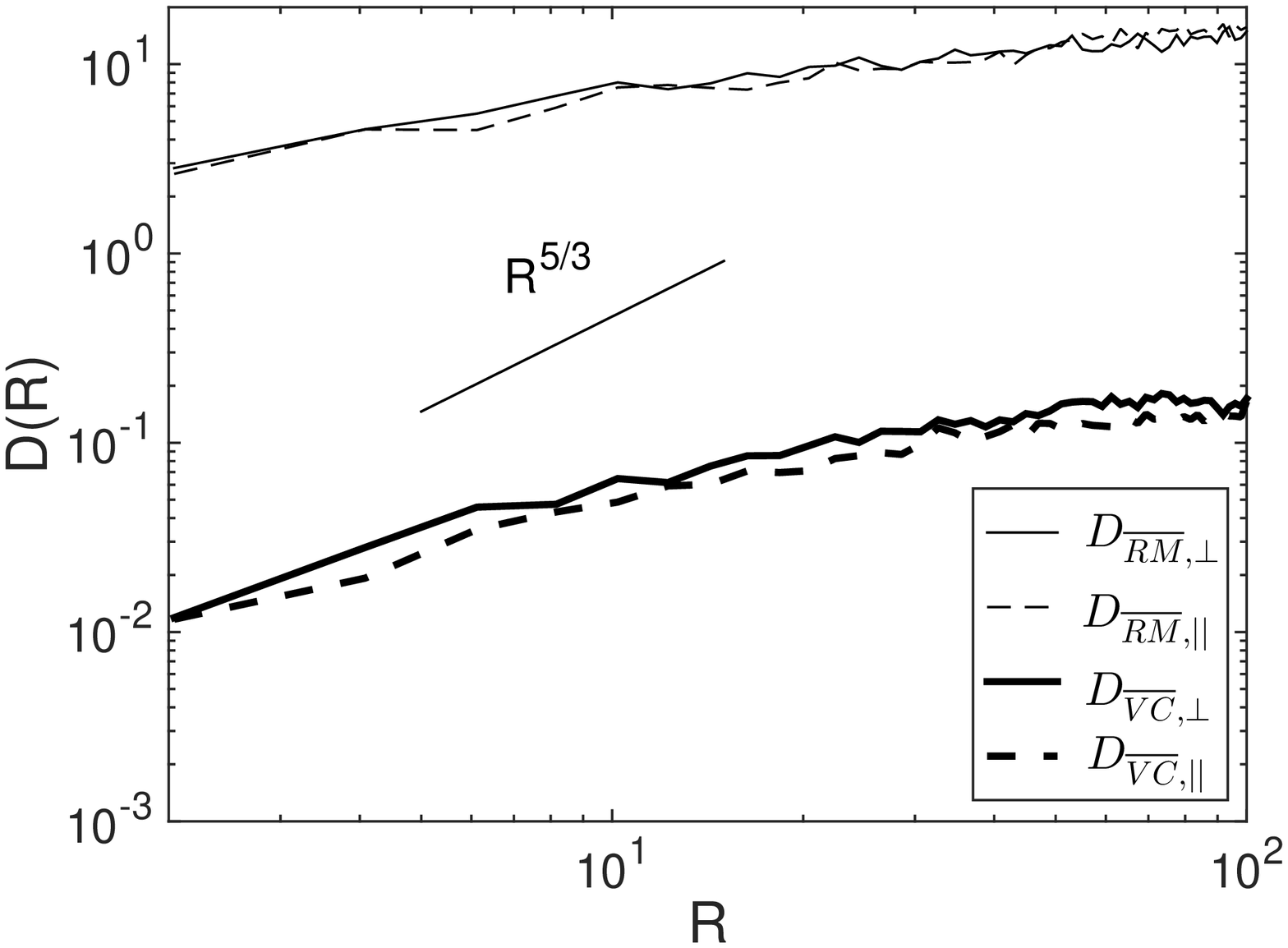}\label{fig: norsup4}}
\subfigure[]{
   \includegraphics[width=8.5cm]{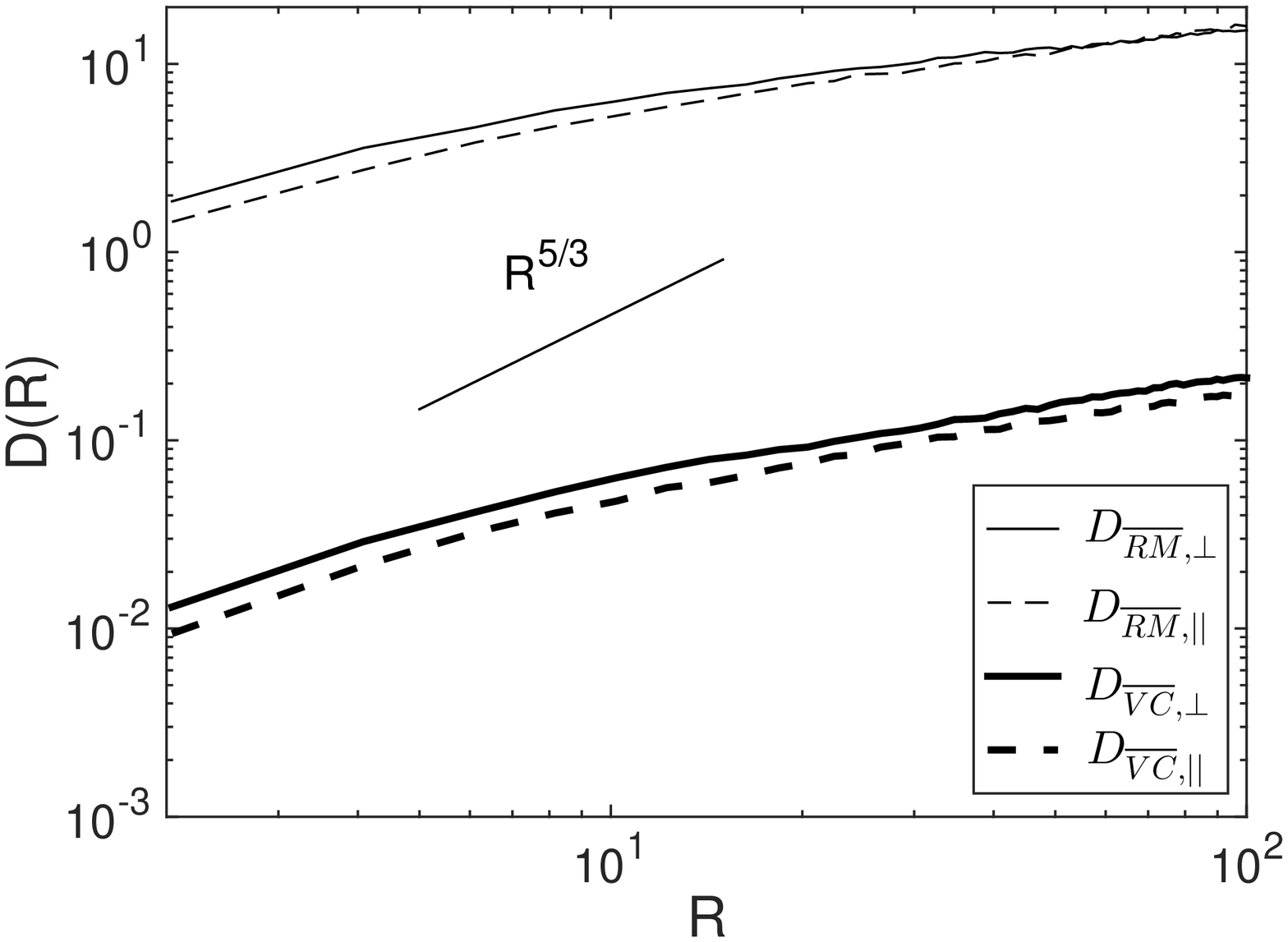}\label{fig: norsup4las}}
\caption{2D SFs for (a) projected magnetic fields, velocities, and densities, (b) RMs and VCs, 
(c) normalized RMs and VCs in supersonic MHD turbulence for run M7. 
The Kolmogorov scaling is indicated by the short solid line.
{(d) Same as (c) but with a larger sample size.}}
\label{fig: sup42d}
\end{figure*}

The anisotropy of 3D SFs was earlier studied by 
\citet{Hua20}. In this work we are interested in the anisotropy of 2D SFs. As an example, the 2D SFs of 
the projected magnetic fields $\int_0^L dz B_z (\bm{X},z)$,
velocities $\int_0^L dz v_z (\bm{X},z)$ (Eq. \eqref{eq: drpjt}), and densities $\int_0^L dz \rho(\bm{X},z)$ for the supersonic MHD turbulence (M7) are presented in Fig.~\ref{fig: colsup4}. 
We use $D_{\perp}$ and $D_{\|}$ to represent $D(R_\perp)$ and $D(R_\|)$ for simplicity. Due to the projection effect, there is $D(R) \propto R^{5/3}$ (Eq. \eqref{eq: app1thick}) for the Kolmogorov scaling. {Due to the solenoidal driving, the Kolmogorov scaling for turbulent velocities is expected even in supersonic MHD turbulence (e.g., \citealt{2007ApJ...658..423K})}. 
$D_\rho$ distinctively exhibits a shallow slope and isotropy compared with $D_b$ and $D_v$.  {Both the shallowness of density spectrum and isotropy of density fluctuations in supersonic MHD turbulence were found in earlier studies, e.g., \citet{BLC05}.} They account for the shallow slope and 
suppressed anisotropy seen for $D_\text{RM}$ and $D_\text{VC}$ in Fig.~\ref{fig: unsup4}. As expected, the fluctuations of RMs and VCs are dominated by density fluctuations in highly supersonic turbulence, and thus 
the anisotropic scalings of turbulent magnetic fields and velocities cannot be fully recovered. For our simulated data, RM and VC are measured as 
\begin{equation}
   \text{RM} = \int_0^Ldz \rho B_z , 
\end{equation}
and 
\begin{equation}
   \text{VC} = \int_0^L dz \rho v_z . 
\end{equation}
The above approximation for RM is valid when the ionization fraction can be treated as a constant for the sampled turbulence volume.

To mitigate the effect of density, we also examine the SFs of normalized RM 
\begin{equation}
    D_{\overline{\text{RM}}} (R) = \Big\langle \Big[\frac{\text{RM}(\bm{X_1})}{\text{DM}(\bm{X_1})} - \frac{\text{RM}(\bm{X_2})}{{\text{DM}(\bm{X_2})}}\Big]^2 \Big\rangle,
\end{equation}
and normalized VC
\begin{equation}\label{eq: vcnosf}
    D_{\overline{\text{VC}}} (R) = \Big\langle \Big[\frac{\text{VC}(\bm{X_1})}{\text{DM}(\bm{X_1})} - \frac{\text{VC}(\bm{X_2})}{{\text{DM}(\bm{X_2})}}\Big]^2 \Big\rangle,
\end{equation}
as shown in Fig.~\ref{fig: norsup4}, where 
\begin{equation}
    \text{DM} = \int_0^L dz \rho 
\end{equation}
is the column density. DM can be treated as dispersion measure $\int_0^L dz n_e$ when the ionization fraction is a constant. We see that the division over DM  slightly improves the performance of RMs and VCs in retrieving the turbulence anisotropy. {To examine the effect of sample size, we also present the measured $D_{\overline{\text{RM}}}$ and $D_{\overline{\text{VC}}}$
with a larger sample size of $10^5$ in Fig.~\ref{fig: norsup4las}. Although increasing the sample size yields a 
smoother SF, the anisotropy is still insignificant.}

\begin{figure*}[htbp]
\centering   
\subfigure[]{
   \includegraphics[width=8.5cm]{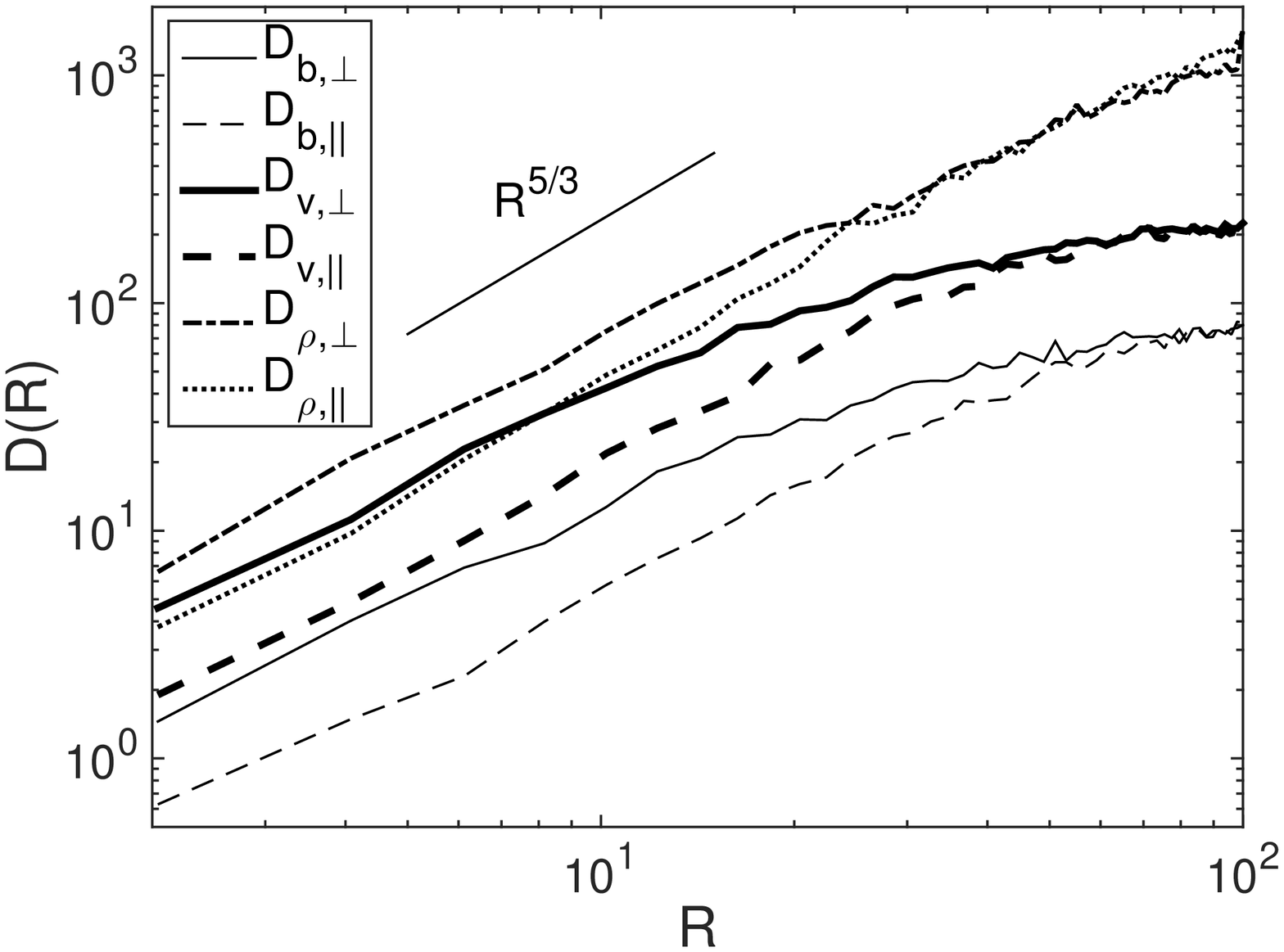}\label{fig: colsub}}
\subfigure[]{
   \includegraphics[width=8.5cm]{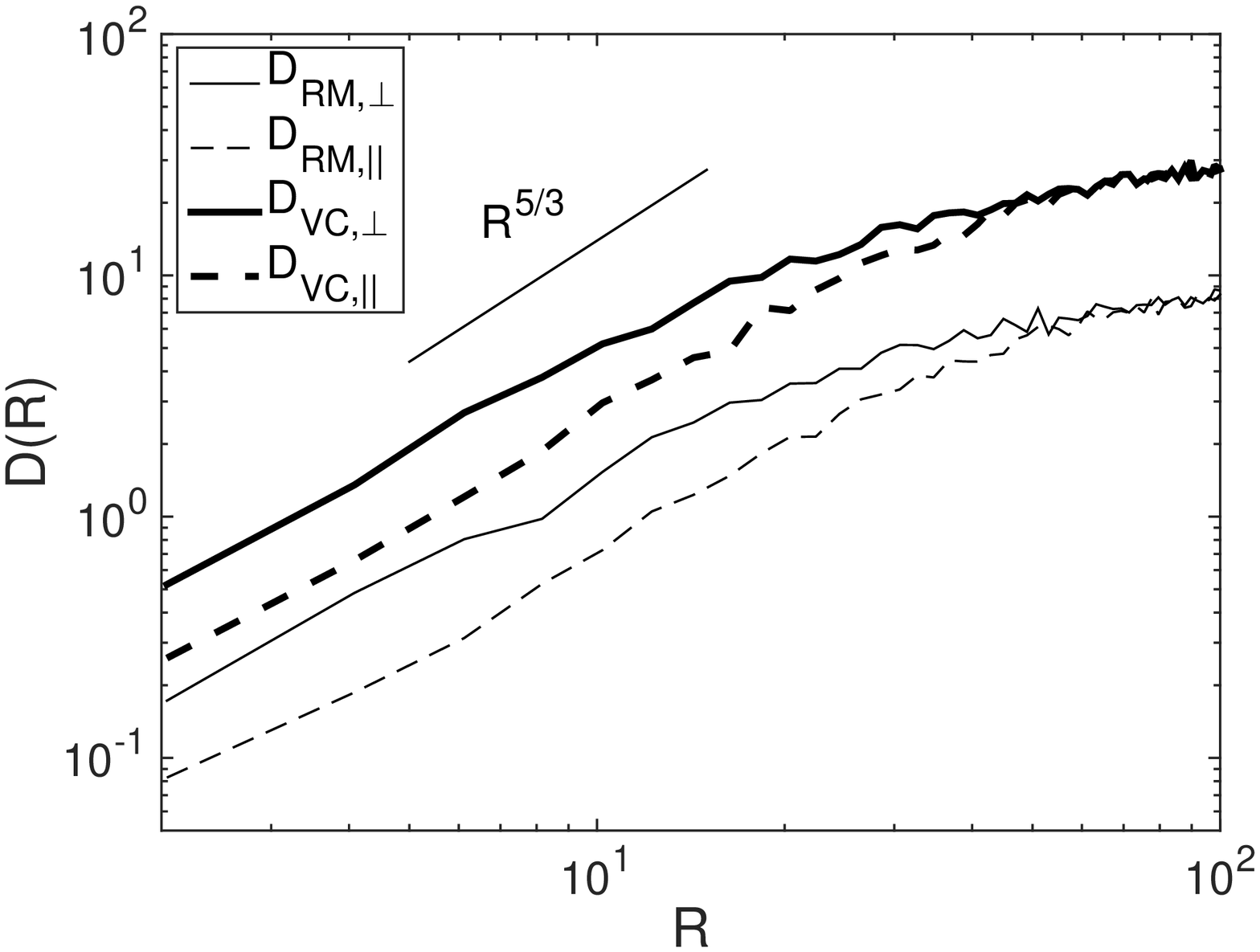}\label{fig: unsub}}
\subfigure[]{
   \includegraphics[width=8.5cm]{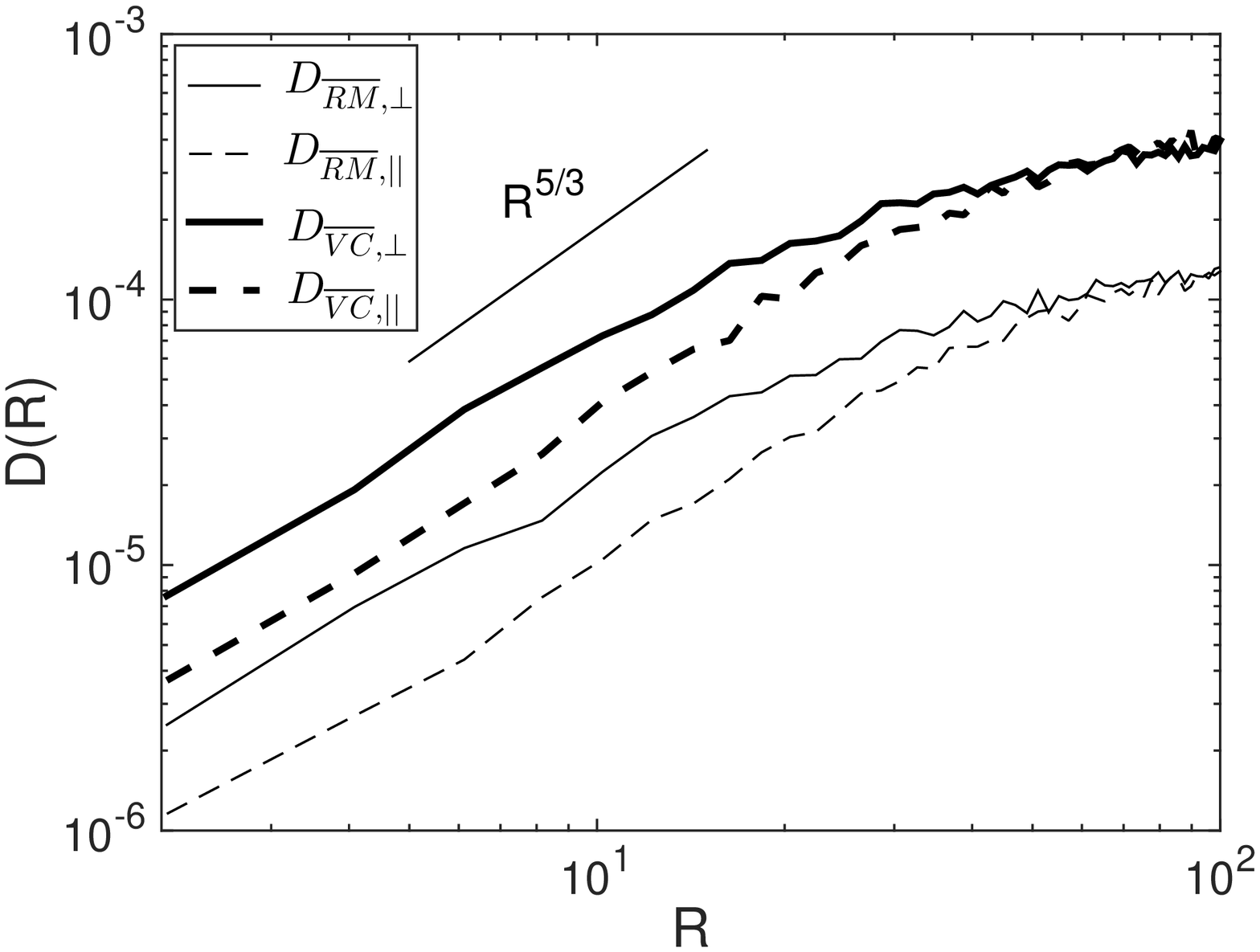}\label{fig: norsub}}
\caption{Same as Fig.~\ref{fig: sup42d} but for subsonic MHD turbulence of run M0. }
\label{fig: subani2dsf}
\end{figure*}

As a comparison, Fig.~\ref{fig: subani2dsf} displays the 2D SFs in subsonic MHD turbulence. For RMs and VCs, the involvement of densities does not affect the anisotropic scalings of magnetic fields and velocities. The applicability of VCs to  subsonic turbulence has been numerically studied by, e.g., \citet{EL05,Esq07,BurL14}. 
The above results suggest that the SF of RMs can also be used to obtain turbulence anisotropy in subsonic MHD turbulence. 

We notice that the effect from isotropic driving and insufficient inertial range is alleviated for 2D SFs. 
The inertial range appears to be more extended due to the projection effect, and the anisotropy on smaller scales becomes scale independent. 

\subsubsection{Low-density regions in supersonic MHD turbulence}
\label{sssec: lowd}
It is clear that the suppressed anisotropy seen from the 
SFs of RMs and VCs in supersonic MHD turbulence is caused by the isotropic density fluctuations. It was earlier found in \citet{BLC05} that by filtering out high-density peaks with the logarithm of density used, the density statistics even in supersonic MHD turbulence exhibit the turbulence anisotropy similar to that in subsonic MHD turbulence. 

\begin{figure*}[htbp]
\centering   
\subfigure[M1]{
   \includegraphics[width=8.5cm]{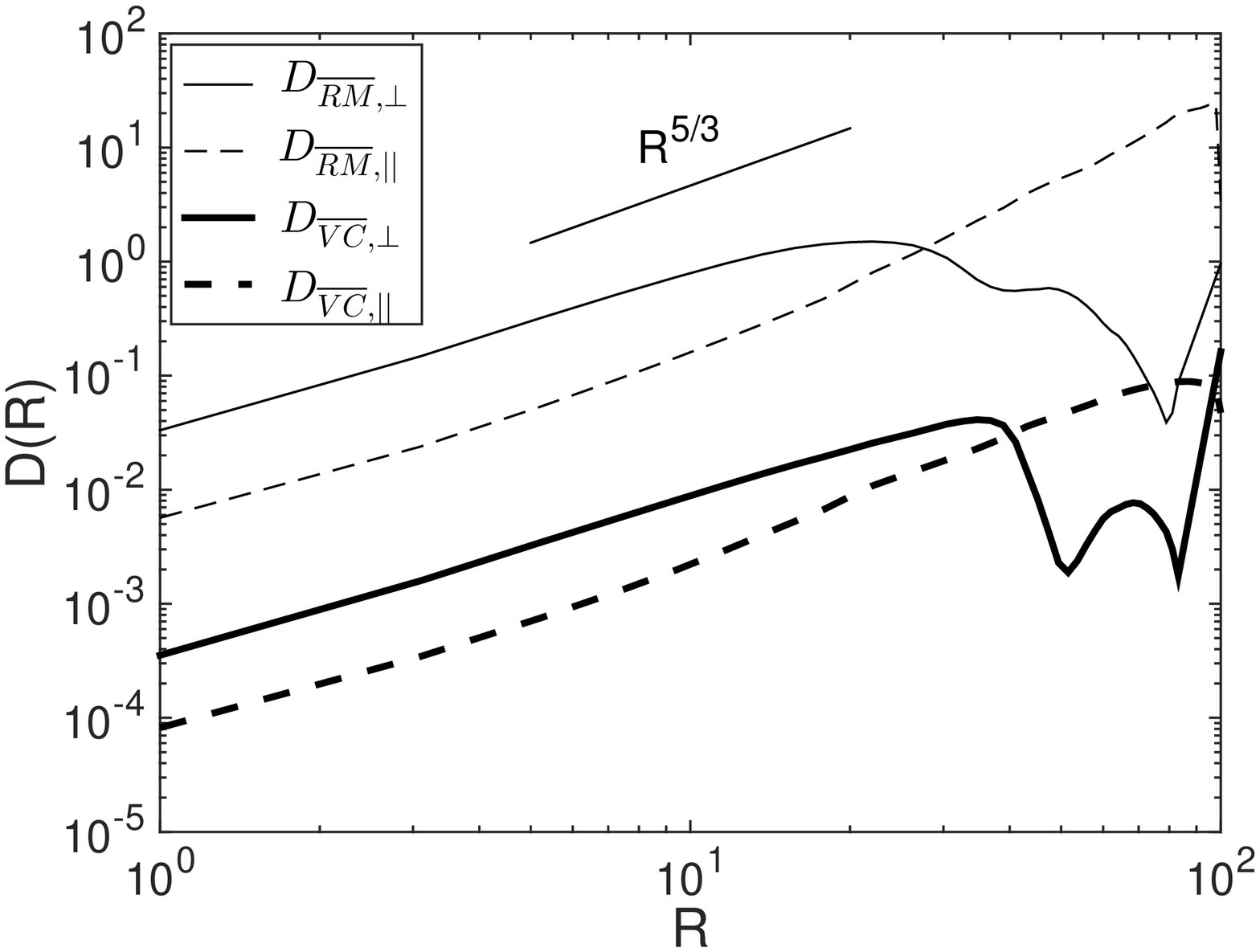}\label{fig: ld0}}
\subfigure[M2]{
   \includegraphics[width=8.5cm]{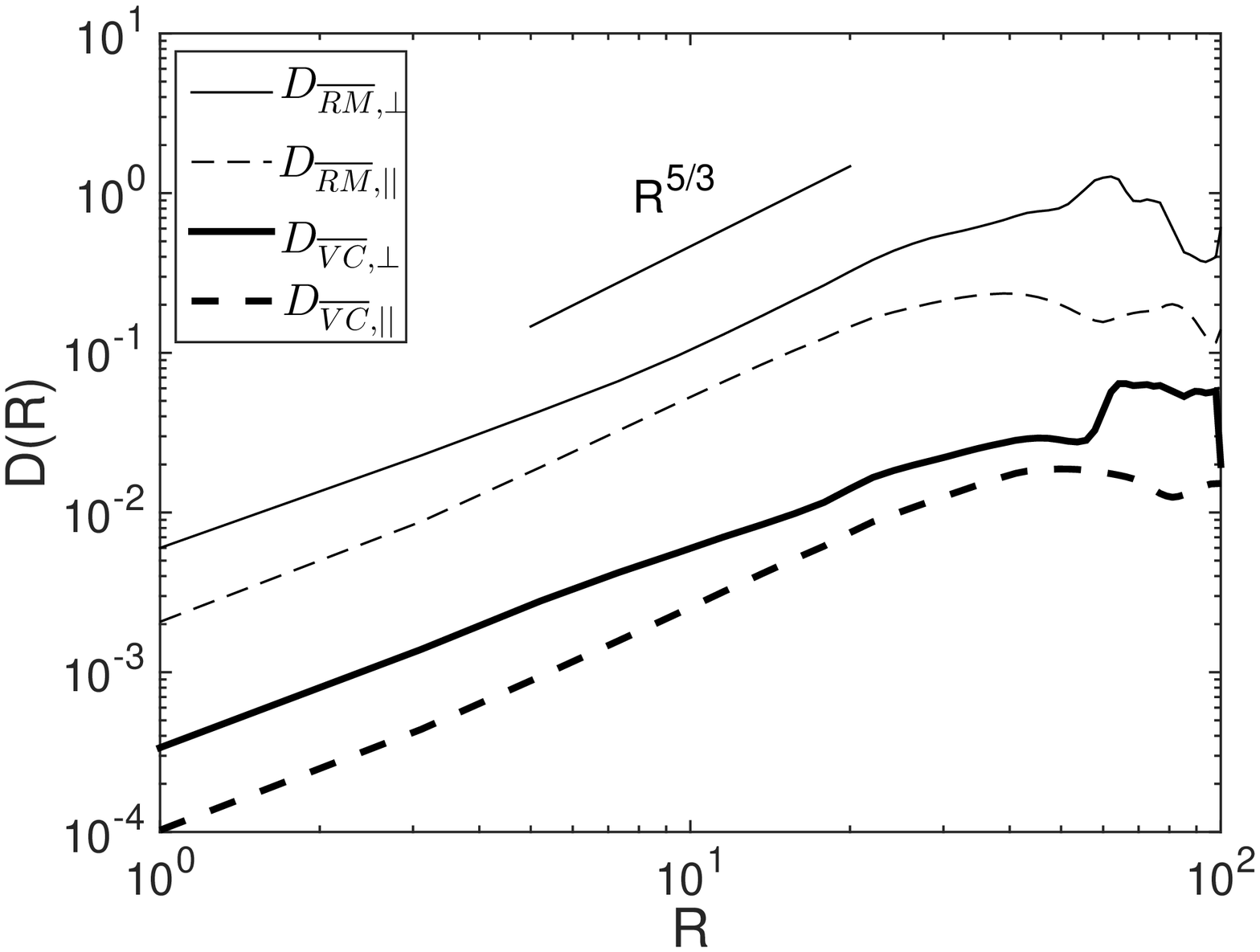}\label{fig: ld1}}
\subfigure[M3]{
   \includegraphics[width=8.5cm]{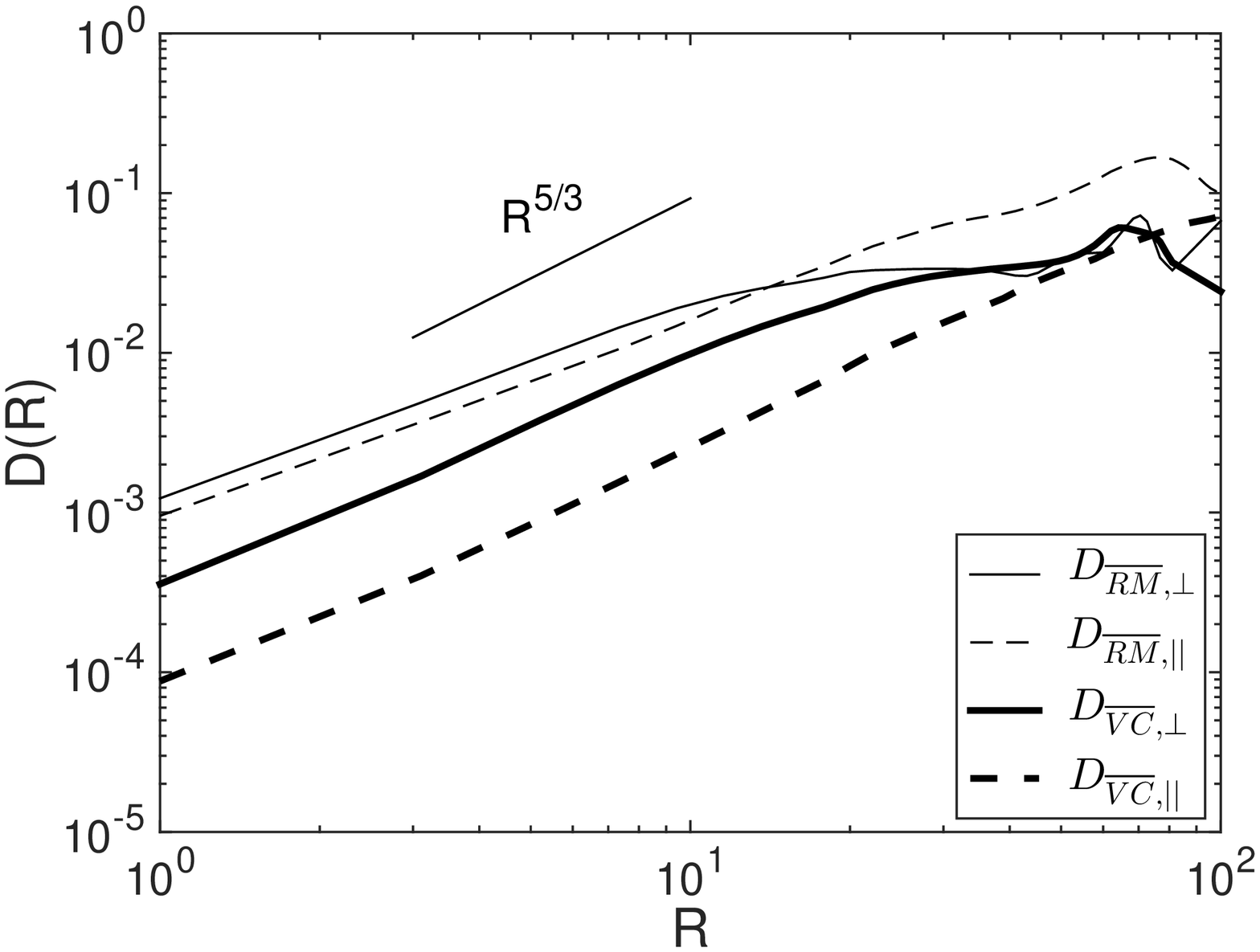}\label{fig: ld2}}
\subfigure[M4]{
   \includegraphics[width=8.5cm]{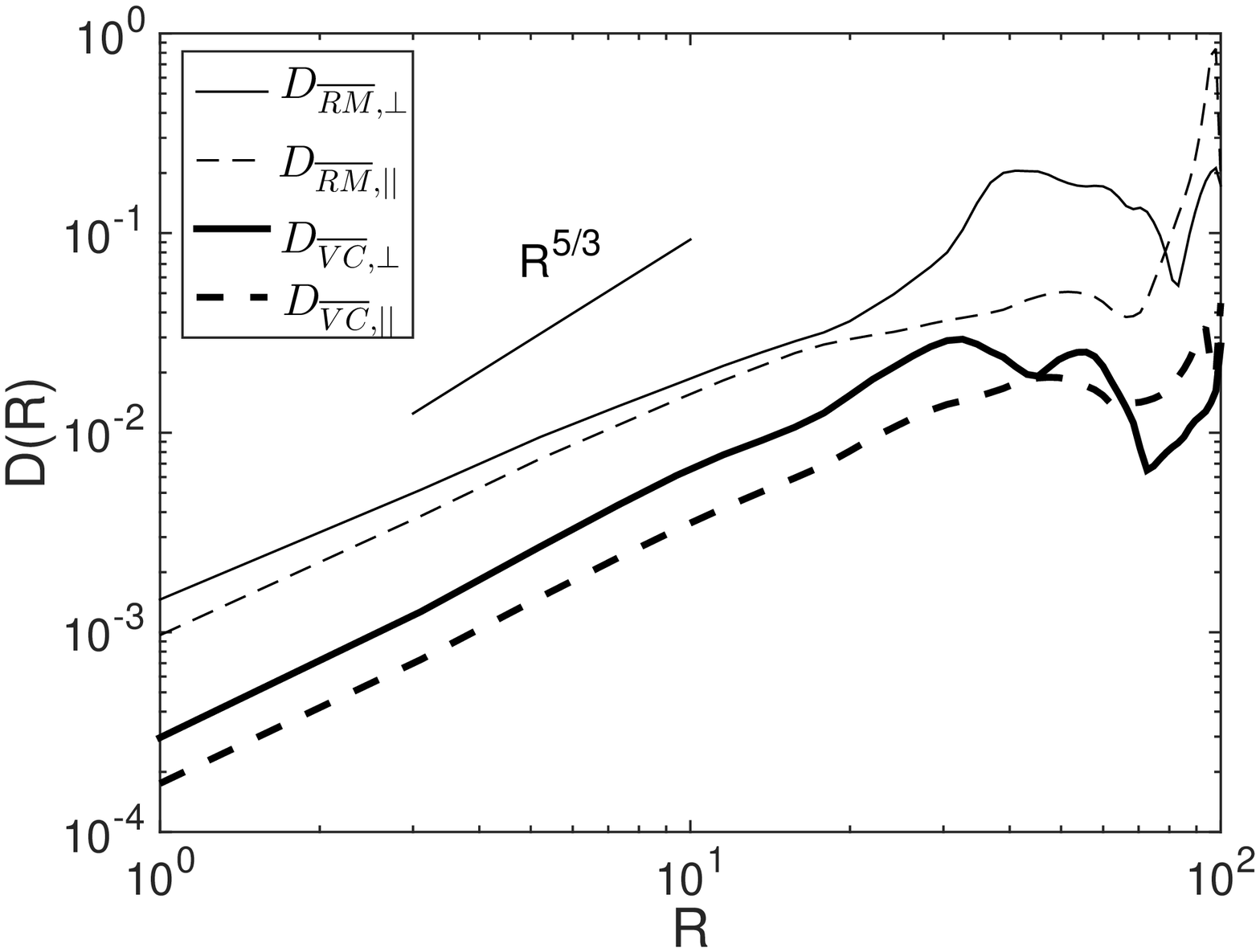}\label{fig: ld3}}
\subfigure[M5]{
   \includegraphics[width=8.5cm]{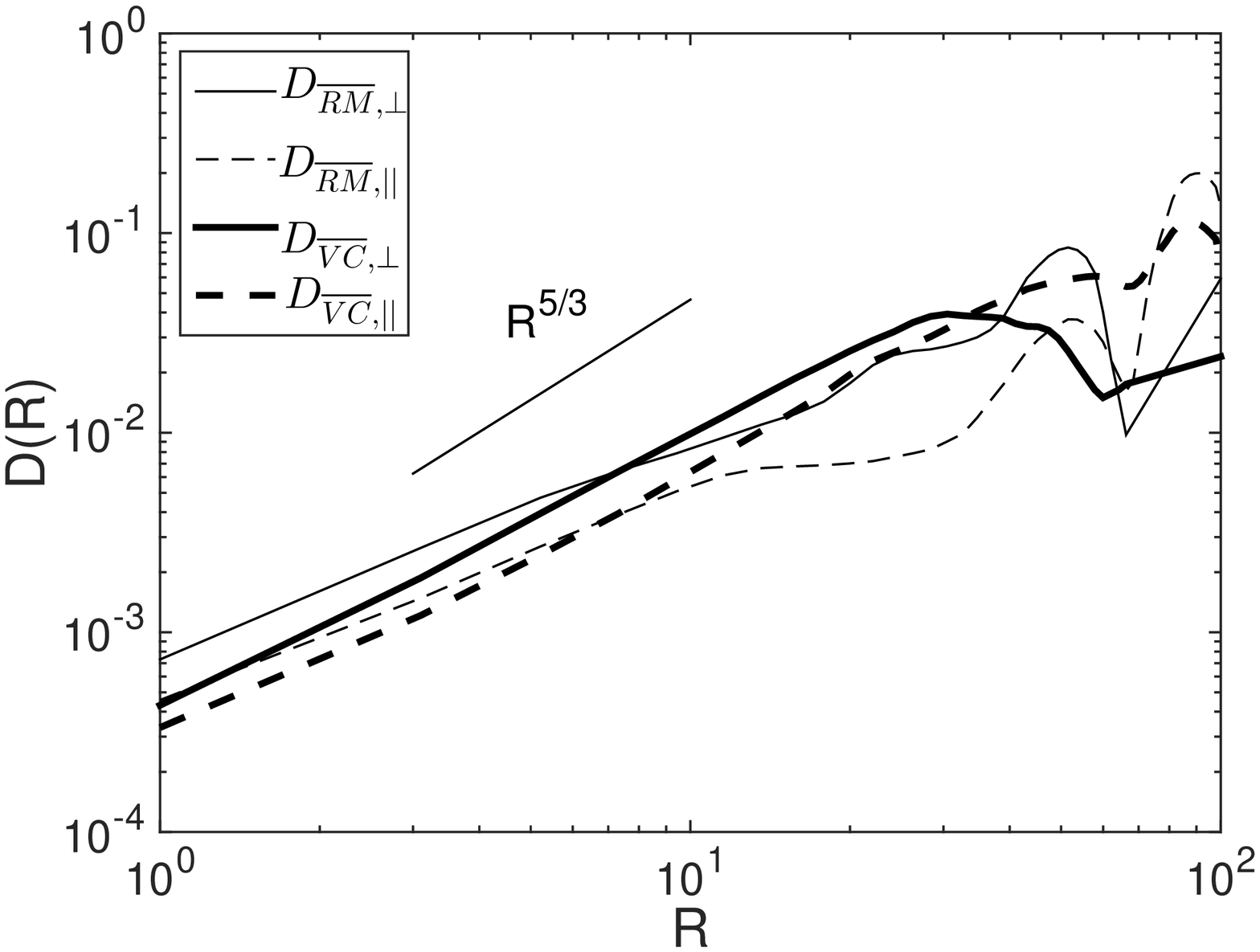}\label{fig: ld4}}
\caption{SFs of selectively sampled
$\overline{\text{RM}}$s and 
$\overline{\text{VC}}$s in low-density regions in supersonic MHD turbulence for 
runs M1-M5 with $M_S \sim 6-7$.}
\label{fig: ldlsup}
\end{figure*}

\begin{figure*}[htbp]
\centering   
\subfigure[M6]{
   \includegraphics[width=8.5cm]{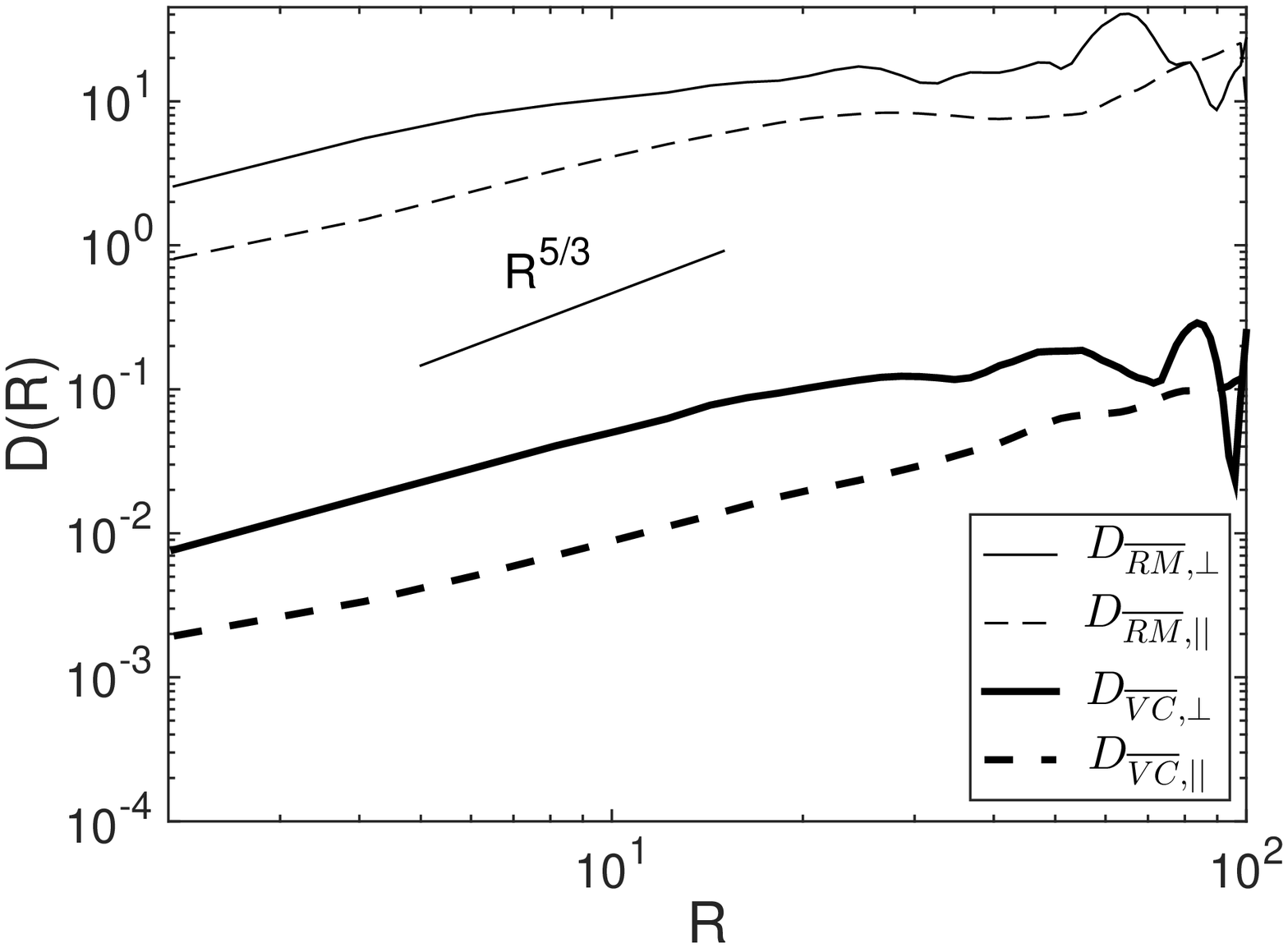}\label{fig: ldsup2}}
\subfigure[M7]{
   \includegraphics[width=8.5cm]{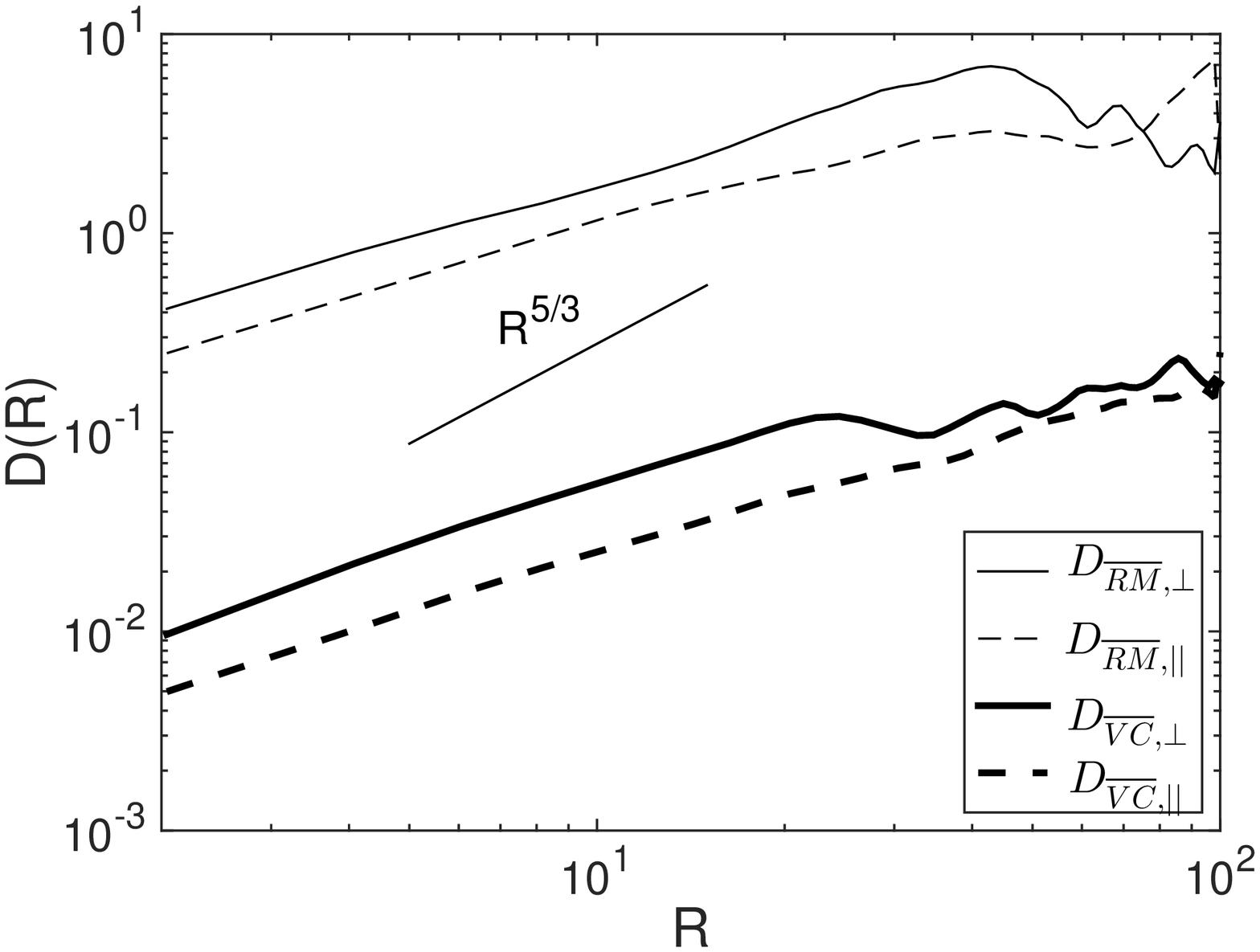}\label{fig: ldsup4}}
\subfigure[M8]{
   \includegraphics[width=8.5cm]{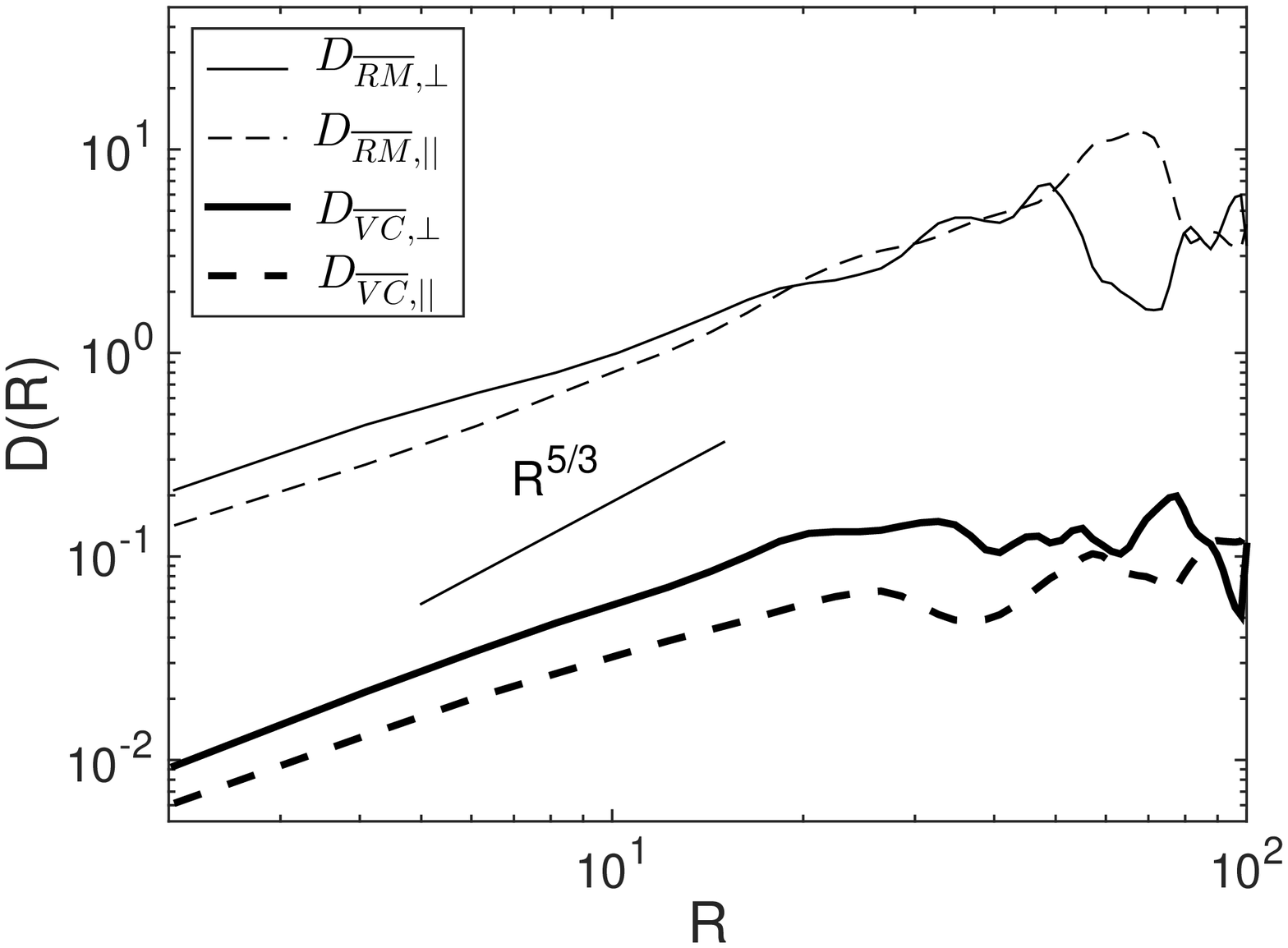}\label{fig: ldsup5}}
\subfigure[M9]{
   \includegraphics[width=8.5cm]{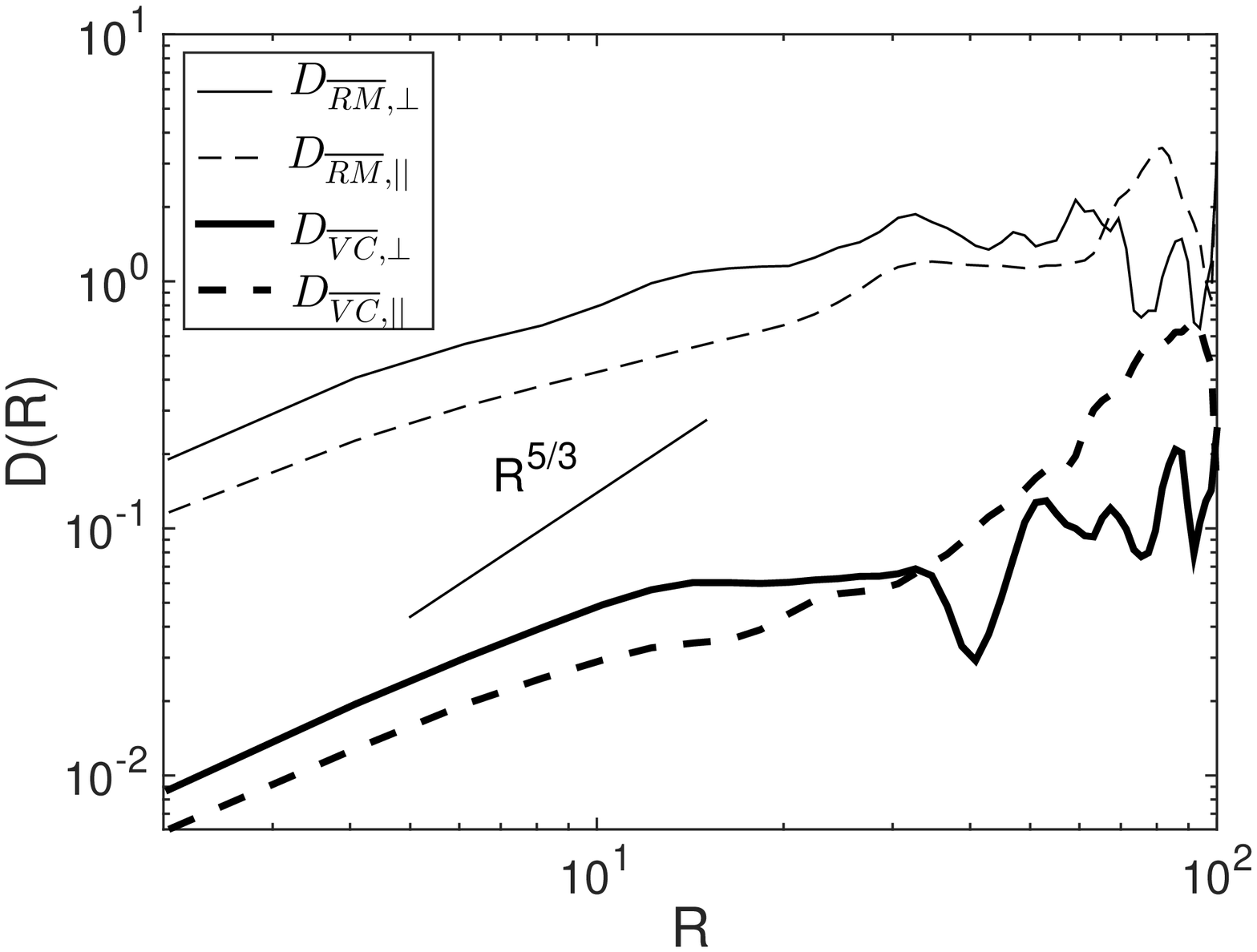}\label{fig: ldsup6}}
\caption{ Same as Fig.~\ref{fig: ldlsup} but for  
runs M6-M9 with $M_S \sim 11$.}
\label{fig: ldhsup}
\end{figure*}

The small-scale density enhancements in supersonic turbulence are spatially clustered and concentrated with a small volume filling factor. The volume-filling densities are mainly in the range $1 / M_S  \lesssim  \rho / \langle \rho \rangle \lesssim 1 $ \citep{RoG18}. To avoid the distortions by high-density peaks, we only select the lines of sight through the relatively low-density regions with 
\begin{equation}
      \text{DM} < \langle \rho \rangle  L .
\end{equation}
The resulting SFs of $\overline{\text{RM}}$s and $\overline{\text{VC}}$s in low-density regions are presented in Figs.~\ref{fig: ldlsup} and \ref{fig: ldhsup} for supersonic MHD turbulence with $M_S \sim 6-7$ and $M_S \sim 11$, respectively (see Table~\ref{tab:num}). Compared with the case using random sampling in Fig.~\ref{fig: sup42d},
the selective sampling adopted here leads to more anisotropic SFs of both $\overline{\text{RM}}$s and $\overline{\text{VC}}$s in supersonic MHD turbulence 
over an extended range of length scales.  {The SFs on large scales get deformed because of our selective sampling method. The incomplete sampling of the data cube causes loss of information on large-scale turbulent fluctuations, but this does not affect the ansitropy measurement on small scales (see below).} We note that the density effect is not fully suppressed, as $D_{\overline{\text{RM}}}(R)$ and $D_{\overline{\text{VC}}}(R)$ still show deviations from the Kolmogorov scaling, especially in the highly supersonic case in Fig.~\ref{fig: ldhsup}.

\begin{figure*}[htbp]
\centering   
\subfigure[Selective sampling of low-density regions]{
   \includegraphics[width=8.5cm]{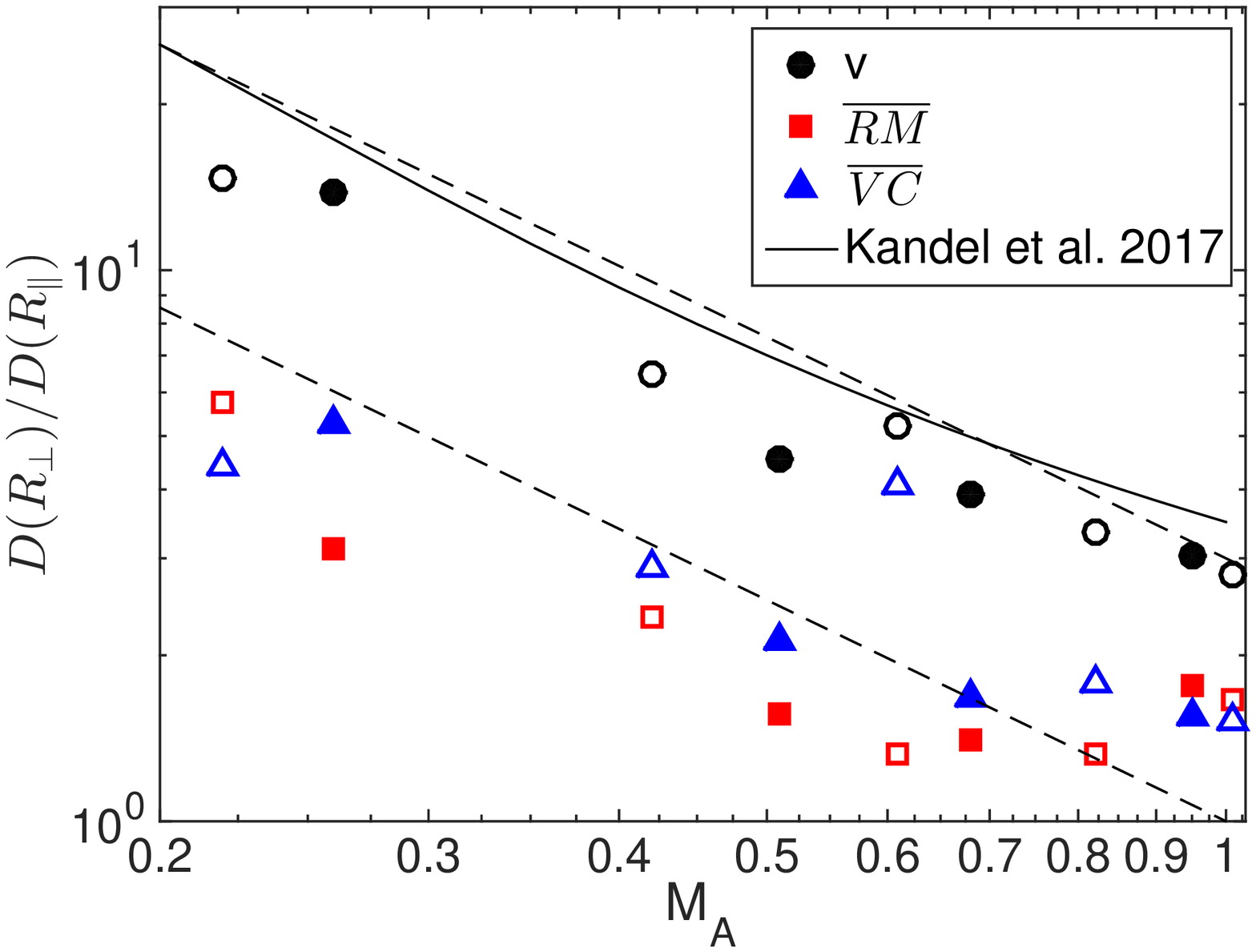}\label{fig: ani}}
\subfigure[Random sampling]{
   \includegraphics[width=8.5cm]{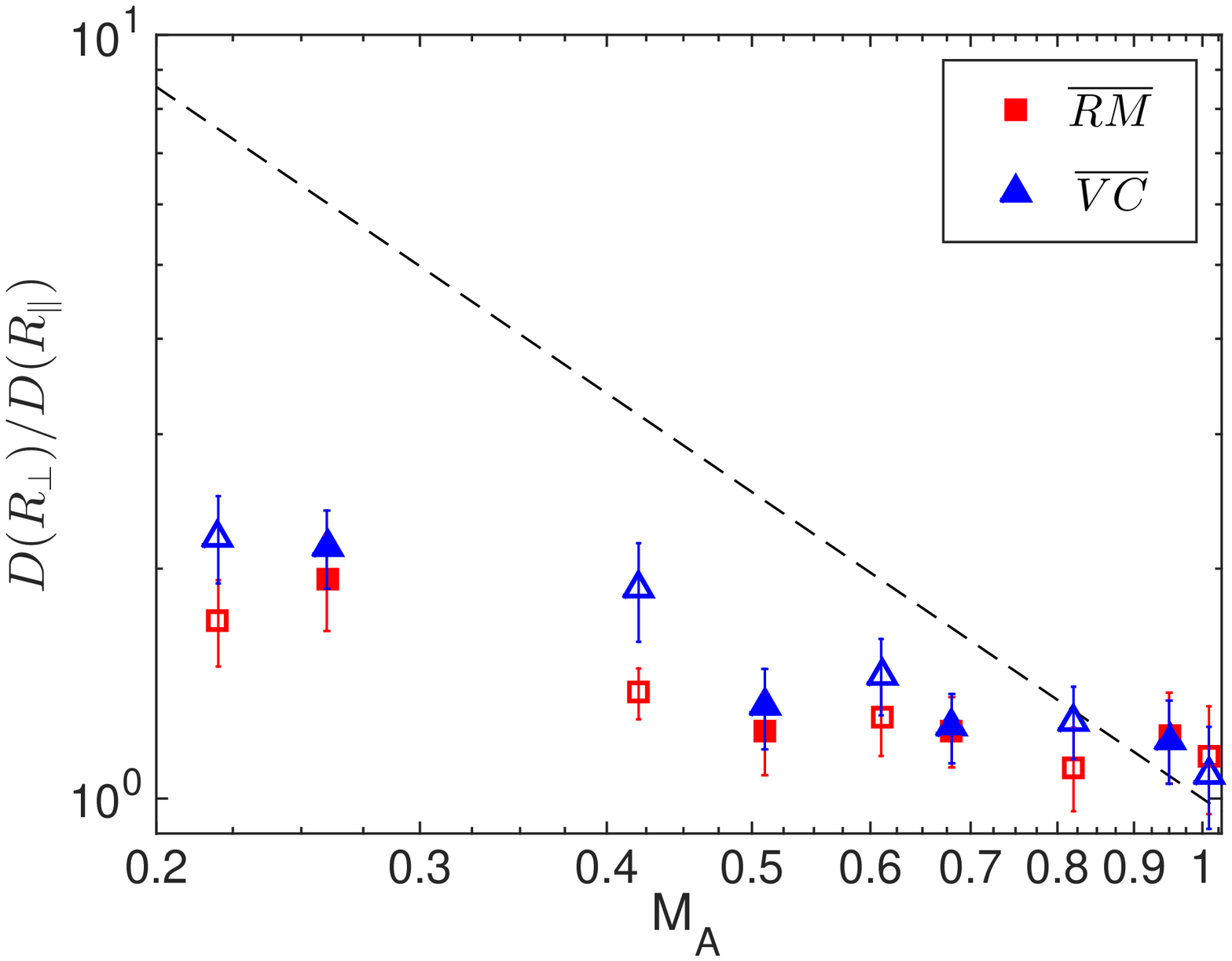}\label{fig: aniran}}
\caption{(a) $D_v(R_\perp) / D_v(R_\|)$, $D_{\overline{\text{RM}}}(R_\perp)/D_{\overline{\text{RM}}}(R_\|)$, and
$D_{\overline{\text{VC}}}(R_\perp)/D_{\overline{\text{VC}}}(R_\|)$ as a function of $\rm M_A$ obtained in low-density regions of supersonic MHD turbulence for runs M1-M9. The open symbols correspond to runs M1-M5, and the filled symbols correspond to runs M6-M9. The dashed lines indicate $\rm M_A^{-4/3}$ and $\rm 3M_A^{-4/3}$. The solid line shows the analytical prediction by \citet{Kan17}. (b) Same as (a) but for $D_{\overline{\text{RM}}}$ and $D_{\overline{\text{VC}}}$ with random sampling. The dashed line corresponds to $\rm M_A^{-4/3}$.}
\end{figure*}

From Figs.~\ref{fig: ldlsup} and \ref{fig: ldhsup}, we easily see that the degree of anisotropy varies in different numerical runs with different $\rm M_A$. To quantify the dependence of the anisotropies of $D_{\overline{\text{RM}}}$ and $D_{\overline{\text{VC}}}$ on $\rm M_A$, 
we measure the ADs, i.e.,  
$D_{\overline{\text{RM}}}(R_\perp)/D_{\overline{\text{RM}}}(R_\|)$ and 
$D_{\overline{\text{VC}}}(R_\perp)/D_{\overline{\text{VC}}}(R_\|)$, 
averaged over a range of length scales from one to ten grid cells,
{where the anisotropy remains constant\footnote{Among our numerical runs, M1 has the smallest $M_A=0.22$ (see Table \ref{tab:num}). The corresponding $L_\text{st} = L_i M_A^2$ (Eq. \eqref{eq: sub}) is around $19$ grid cells. The range of length scales for measuring AD is still in the strong turbulence regime.}.
The SFs on large scales are subject to the effects of isotropic driving
\citep{CV00,2018ApJ...865...54Y}
and incomplete sampling in this case, so they are not used for the AD measurement.}
The dependence of AD on $\rm M_A$ is displayed in Fig.~\ref{fig: ani}.
In most cases, the AD seen from $D_{\overline{\text{RM}}}$ is slightly smaller than that from $D_{\overline{\text{VC}}}$.
This is expected, as magnetic fluctuations exhibit a smaller anisotropy compared with velocity fluctuations in supersonic MHD turbulence, 
as shown from their 3D and 2D SFs (Figs. \ref{fig: sup043d} and \ref{fig: colsup4}). The numerically measured $\rm M_A$ dependence is consistent with $\rm M_A^{-4/3}$, which we analytically derived for the
Kolmogorov scaling of turbulence in Section \ref{sssec:2dsf}.
{It shows that despite the deviations from the Kolmogorov slope that are still seen in $D_{\overline{\text{RM}}}$ and $D_{\overline{\text{VC}}}$ measured in low-density regions in supersonic MHD turbulence, the theoretically expected $\rm M_A$ 
dependence of turbulence anisotropy can be recovered.} Note that the non-unity AD at $\rm M_A \approx 1$ is caused by the presence of uniform magnetic field of our initial setup. The dependence on $\rm M_S$ is insignificant.

\begin{figure}[htbp]
\centering   
\includegraphics[width=8.5cm]{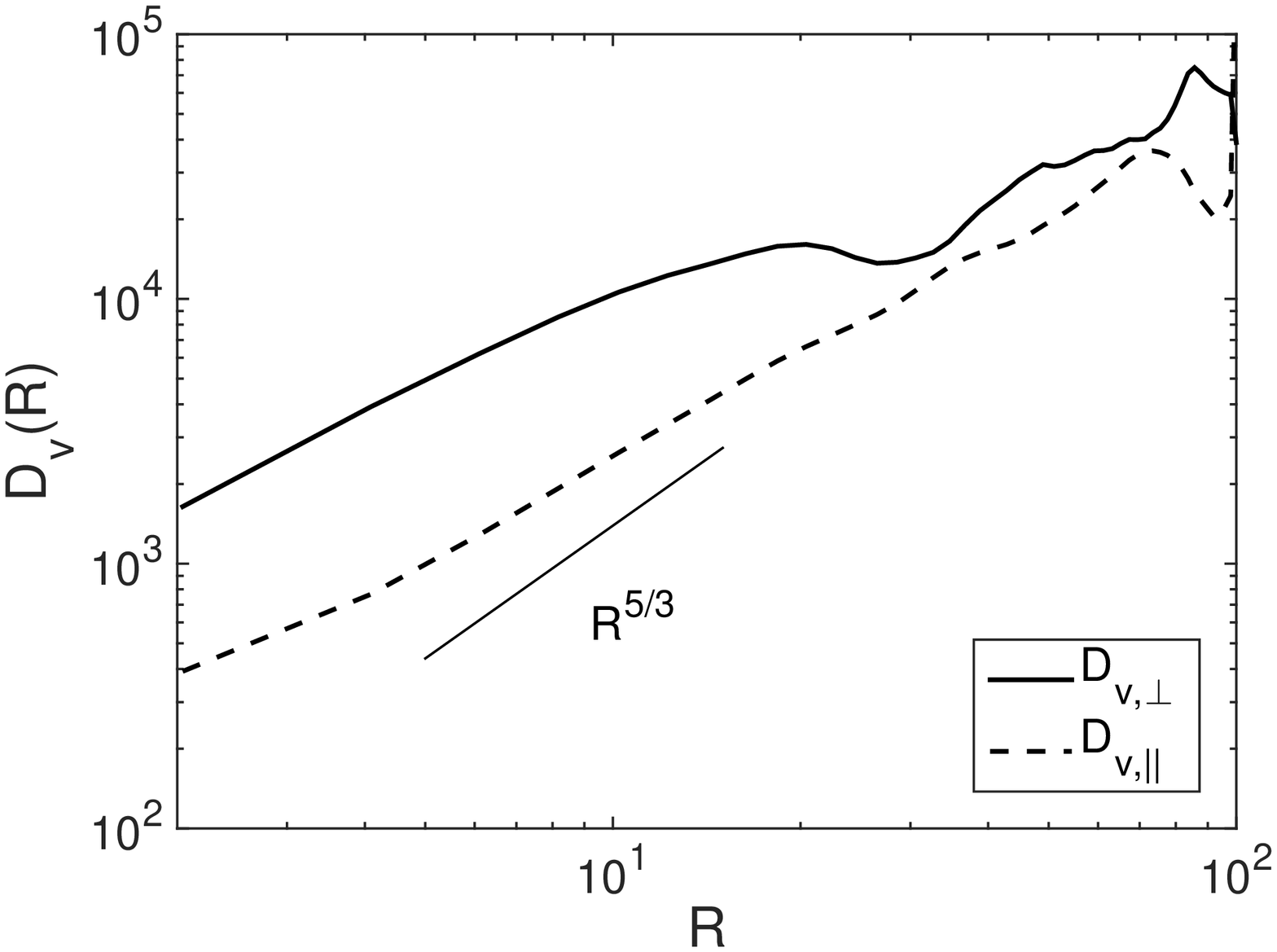}
\caption{SFs of projected velocities in low-density regions of supersonic MHD turbulence for run M7. }
\label{fig: sup4vel}
\end{figure}

To further examine the effect of density on the measured anisotropies of $D_{\overline{\text{RM}}}$ and $D_{\overline{\text{VC}}}$, 
we also measure the ratio $D_v(R_\perp) / D_v(R_\|)$ in low-density regions in supersonic MHD turbulence (see Fig.~\ref{fig: sup4vel} as an example) at different $\rm M_A$. Obviously, the AD obtained from projected turbulent velocities alone is significant larger than those obtained from $D_{\overline{\text{RM}}}$ and $D_{\overline{\text{VC}}}$ at the same $\rm M_A$ (see Fig.~\ref{fig: ani}). Compared with our simple scaling argument in Section~\ref{sssec:2dsf}, it has the $\rm M_A$-dependence better described by the more rigorous theoretical result in \citet{Kan17}, which can be numerically approximated by $\rm 3 M_A^{-4/3}$. It shows that in low-density regions the involvement of densities in $D_{\overline{\text{RM}}}$ and $D_{\overline{\text{VC}}}$ 
affects the observed AD, but not its dependence on $\rm M_A$. 
Since usually $D_v$ is not directly accessible to observations, 
our numerical result can be used to estimate $\rm M_A$ with $\overline{\text{RM}}$s and $\overline{\text{VC}}$s measured in low-density regions of supersonic turbulence in cold interstellar phases. 

As a comparison, we also present $D_{\overline{\text{RM}}}(R_\perp)/D_{\overline{\text{RM}}}(R_\|)$ and 
$D_{\overline{\text{VC}}}(R_\perp)/D_{\overline{\text{VC}}}(R_\|)$
measured with random sampling in supersonic MHD turbulence at different $\rm M_A$ in Fig.~\ref{fig: aniran}. 
Compared with the measurements in low-density regions, 
the ADs are considerably smaller and have a much weaker dependence on $\rm M_A$. {We use the standard errors of the bin averages to calculate the uncertainties in SF measurements related to the sample size. The error of the ratio $D(R_\perp)/D(R_\|)$ is further determined with error propagation, represented by the error bars in Fig.~\ref{fig: aniran}. We note that the error bars are not shown in Fig.~\ref{fig: ani} as they are smaller than the symbol size. For the selectively sampled low-density regions in smaller sub-volumes of the data cube, 
the number of pairs at small separations is larger than that with random sampling, leading to smaller uncertainties.} 

\subsection{Synthetic observations of VCs}
\label{ssec:synobs}
To test the applicability of $\overline{\text{VC}}$s to measuring the magnetization in cold interstellar phases, we produce synthetic observations of CO emission by performing radiative transfer calculations based on our supersonic MHD simulations M1-M9.  The synthetic emission lines of CO isotopologs, i.e., $\rm ^{12}CO$ (1-0) and $\rm C^{18}O$ (1-0), are generated through the SPARX radiative transfer code \citep{2019ApJ...873...16H}. The SPARX solves the radiative transfer equation through the Accelerated Lambda Iteration (ALI). The radiative interaction accounts for the  molecular spontaneous emission, stimulated emission, and the collision with gas particles in the Local Thermodynamic Equilibrium (LTE) condition. Following \citet{2019ApJ...873...16H}, the fractional abundance of the CO isotopolog $\rm ^{12}CO$ (1-0) is set to $1\times10^{-4}$, which comes from the cosmic value of C/H = 3$\times10^{-4}$ and the assumption that 15\% of C is in the molecular form. By adopting $^{12}$CO/C$^{18}$O = 588, we have the fractional abundance of $\rm C^{18}O$ (1-0) as $1.7\times10^{-7}$.

The information about molecular gas density and velocity is taken from our MHD simulations (see Section \ref{sec:data}).
For each scale-free MHD simulation, we choose a cube size of $10$ pc, a gas temperature of $10$ K {(which corresponds to sound speed $c_s$ = 187 m s$^{-1}$)}, and a mean mass density of $300$ $g$ $\rm cm^{-3}$. 
{The initial magnetic field strength is then determined based on the $\rm M_A$ value (see Table \ref{tab:num}).}
The mean optical depth, which mainly depends on the molecular abundance and mean density, is approximately 200 for optically thick $\rm ^{12}CO$ (1-0) and 0.33 for optically thin $\rm C^{18}O$ (1-0), {which is close to the observational value \citep{2016ApJ...826..193L}}.

\begin{figure*}[htbp]
\centering   
\subfigure[$^{12}$CO]{
   \includegraphics[width=8cm]{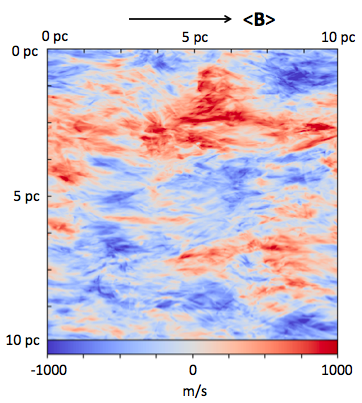}\label{fig: 12map}}
\subfigure[C$^{18}$O]{
   \includegraphics[width=8cm]{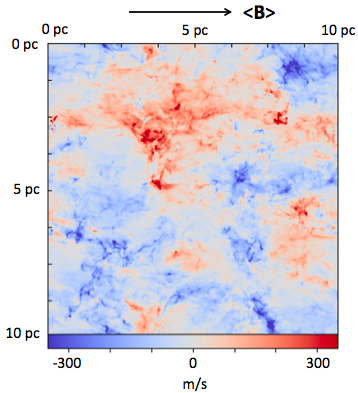}\label{fig: 18map}}
\caption{$\overline{\text{VC}}$ maps of (a) $^{12}$CO and (b) C$^{18}$O emission for run M7.
Note that the color scales in (a) and (b) are different.}
\label{fig:vcmap}
\end{figure*}

The synthetic $\overline{\text{VC}}$ map is calculated using:
\begin{equation}
    \overline{\text{VC}}=\frac{\int T_{R}\cdot v_z dv_z}{\int T_{R} dv_z},
\end{equation}
where $T_R$ is the radiation temperature, and $v_z$ is the LOS velocity.  We note that for optically thin emission lines, the above definition of $\overline{\text{VC}}$ is equivalent to that used in Eq. \eqref{eq: vcnosf} \citep{Kan17}. As an example, the $\overline{\text{VC}}$ maps of $^{12}$CO and C$^{18}$O for run M7 are displayed in Fig.~\ref{fig:vcmap}. The small-scale dense regions with small velocity fluctuations are preferentially traced by C$^{18}$O, 
while the relatively diffuse gas on large scales with large velocity fluctuations is primarily traced by $^{12}$CO, resulting in larger $\overline{\text{VC}}$s of $^{12}$CO. In addition, we can easily see that the structures in the $^{12}$CO map are more anisotropic than those in the C$^{18}$O map, which are aligned with the mean magnetic field.

\begin{figure*}[htbp]
\centering   
\subfigure[]{
   \includegraphics[width=8.5cm]{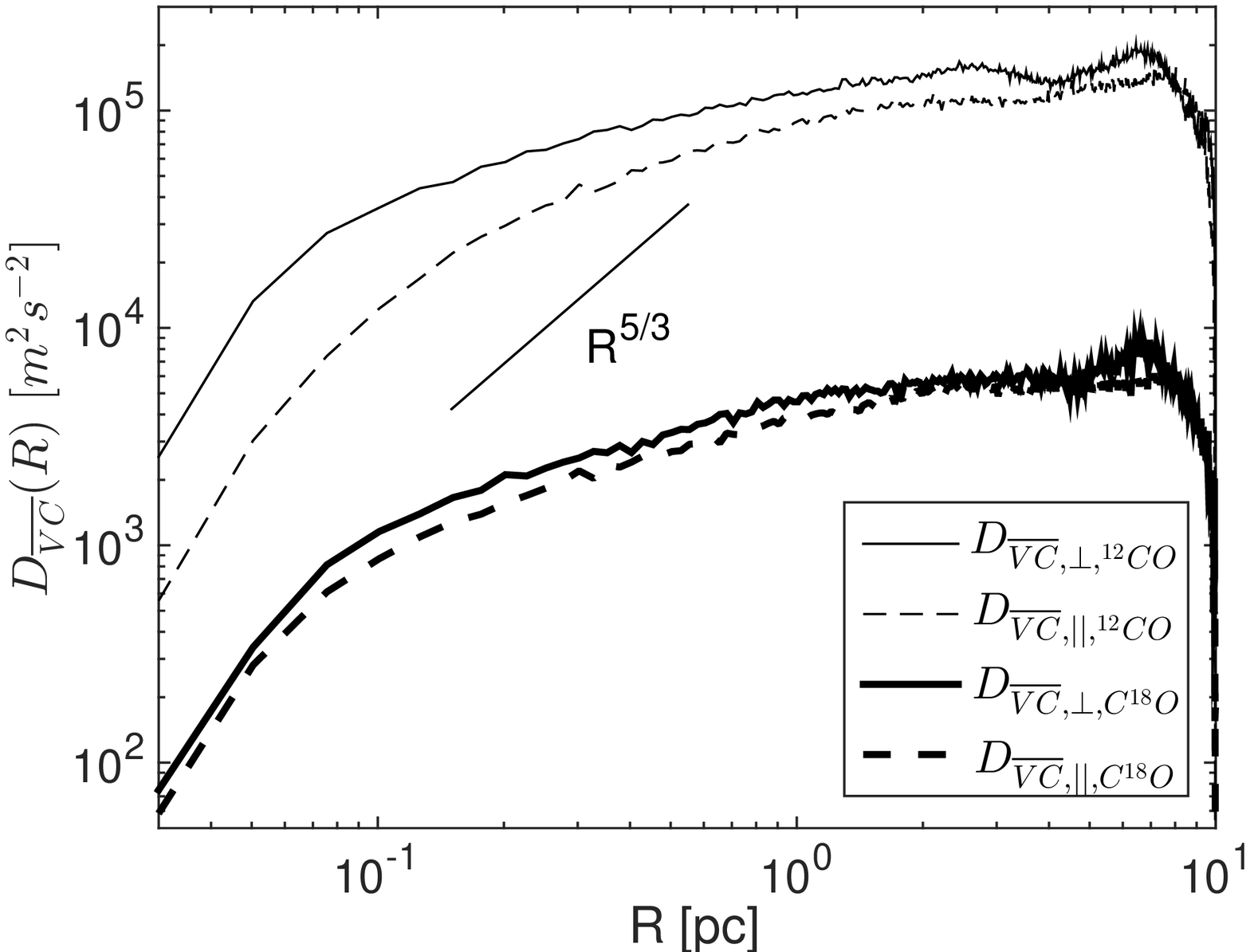}\label{fig: cosf}}
\subfigure[]{
   \includegraphics[width=8.5cm]{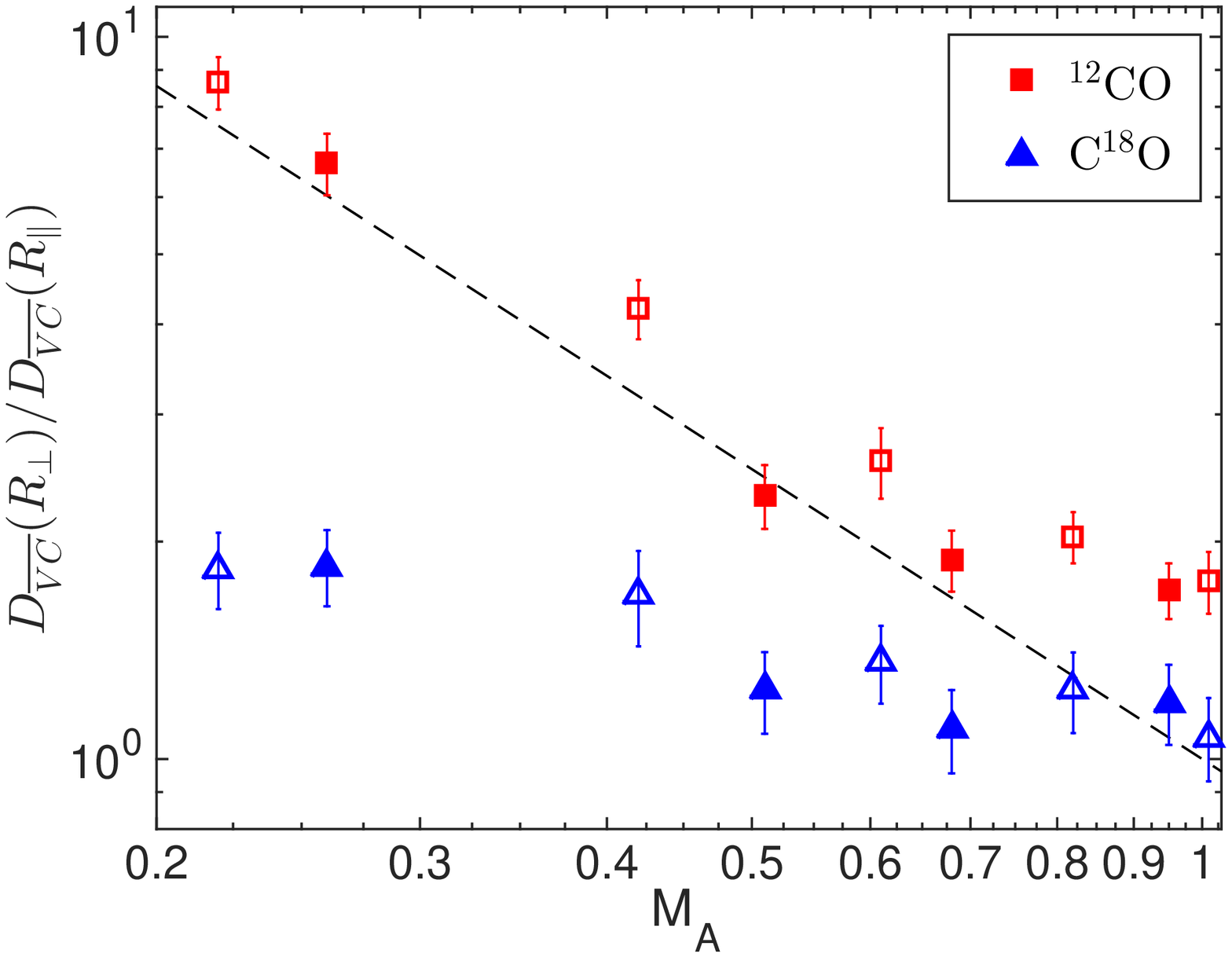}\label{fig: coani}}
\caption{(a) SFs of 
$\overline{\text{VC}}$s of 
$^{12}$CO and C$^{18}$O
for run M7.
(b) $D_{\overline{\text{VC}}}(R_\perp)/D_{\overline{\text{VC}}}(R_\|)$ as a function of $\rm M_A$ measured with 
$^{12}$CO and C$^{18}$O. 
The open symbols correspond to runs M1-M5, and the filled symbols correspond to runs M6-M9. 
The dashed line shows $\rm M_A^{-4/3}$.  }
\end{figure*}

For the quantitative anisotropy analysis, we measure the $\overline{\text{VC}}$ SFs of both $^{12}$CO and C$^{18}$O in the directions perpendicular and parallel to the mean magnetic field, as shown in Fig.~\ref{fig: cosf} as an example.  Different from the 2D SFs without radiative transfer, the $\overline{\text{VC}}$ SFs of $^{12}$CO and C$^{18}$O steepen toward smaller scales, due to the stronger self-absorption in small-scale dense regions. This is different from the case with homogeneous density distribution,  where the self-absorption effect becomes more important toward larger LOS separations, and the turbulence scaling can only be seen
on sufficiently small scales \citep{Kan17}. Obviously, the difference between $D_{\overline{\text{VC}}}(R_\perp)$ and $D_{\overline{\text{VC}}}(R_\|)$ for $^{12}$CO is significantly larger than that for C$^{18}$O. The AD is measured as the ratio $D_{\overline{\text{VC}}}(R_\perp)/D_{\overline{\text{VC}}}(R_\|)$
at around ten grid cells, i.e., $0.1$ pc, where both the effects from isotropic driving and self-absorption are insignificant. {In realistic situations with a more extended inertial range of turbulence, one should measure the AD over a range of length scales where the anisotropy remains constant.} The dependence of AD on $\rm M_A$ is presented in Fig.~\ref{fig: coani}. {The error bars correspond to the standard errors with error propagation (see Section \ref{sssec: lowd}).} When using the low-density tracer $^{12}$CO, we obtain the result similar to that with selective sampling of low-density regions in Section \ref{sssec: lowd} (see Fig.~\ref{fig: ani}). 
$D_{\overline{\text{VC}}}$ recovers the theoretically expected dependence of anisotropy on $\rm M_A$ as $\rm M_A^{-4/3}$. However, when the higher-density tracer C$^{18}$O is used, the anisotropy is suppressed and has a much weaker dependence on $\rm M_A$. This is similar to the result with random sampling in Section \ref{sssec: com} (see Fig.~\ref{fig: aniran}). The above tests with synthetic observations demonstrate that $\overline{\text{VC}}s$ of both $^{12}$CO and C$^{18}$O can be used to measure the orientation of the plane-of-sky magnetic field, and the former can also be used to determine its strength. 

\begin{figure}[htbp]
\centering 
\includegraphics[width=10cm]{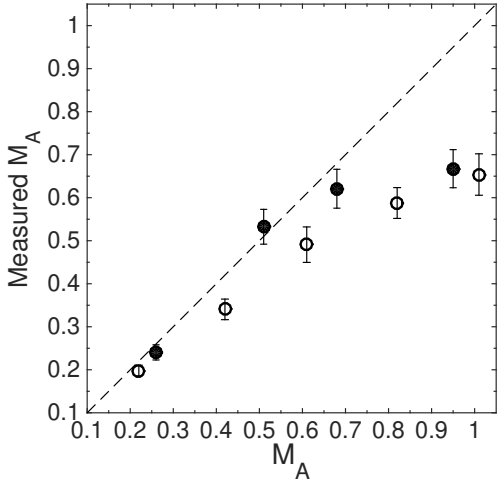}
\caption{$\rm M_A$ measured with $^{12}$CO vs. the real value. The open circles correspond to runs M1-M5, and the filled circles correspond to runs M6-M9. The dashed line represents the equality between the measured and real values.}
\label{fig: mavsma}
\end{figure}

{We further display the $\rm M_A$ values obtained from the ADs measured with $^{12}$CO, i.e., AD$^{-3/4}$, in Fig.~\ref{fig: mavsma}. 
They agree well with the real values for small $\rm M_A$. The discrepancy seen at $\rm M_A\sim 1$ is due to the presence of initial 
uniform magnetic field in our simulations as mentioned before. We also note that for runs M1-M5 with smaller $\rm M_S$ values than those of M6-M9, the measured ADs with $^{12}$CO tend to be larger than the theoretical expectation, leading to the underestimate of $\rm M_A$. The additional uncertainties due to the $\rm M_S$ dependence of AD should be taken into account when measuring $\rm M_A$ with $^{12}$CO.}

\section{Discussion}
\label{sec:4}
{Note that even in low-density regions in supersonic MHD turbulence, the density effect cannot be eliminated from the SFs 
of $\overline{\text{RM}}$s and $\overline{\text{VC}}$s. 
As a result, the SFs have shallower slopes than the Kolmogorov one 
and smaller anisotropies than that of turbulent velocities. In addition, as argued in \citet{Lee16}, it is not easy to reveal the true spectral slope of turbulence by using SFs even with a high numerical resolution. However, our results suggest that the $\rm M_A$ dependence of anisotropy is not very sensitive to the exact slope of SF.}

{When dealing with real observations, the orientation of the POS mean magnetic field is unknown. One should measure the SFs of $\overline{\text{RM}}$s or $\overline{\text{VC}}$s at different position angles on the POS. The direction corresponding to the minimum SF is the direction of the POS mean magnetic field. Then the anisotropy can be measured by using the ratio between the maximum and minimum SFs (measured in directions orthogonal to each other) averaged over a range of length scales with a constant anisotropy. The anisotropy of our interest originates from the magnetization of turbulence, but the observed anisotropy can be affected by various effects. Besides the driving effect and density effect, in dense regions of molecular clouds, both the scaling and anisotropy of turbulence can be affected by gravity. Once the gas traced by $\rm ^{12}CO$ undergoes gravitational collapse, 
the dependence of AD on $\rm M_A$ would get changed.} 

In this work we only considered the RMs and VCs of a turbulence layer with a given thickness, which is applicable when studying, e.g., 
the interstellar RMs for extragalactic radio sources, the VCs of a molecular cloud. In the case of RMs of Galactic pulsars with different distances from the observer, there are extra terms in their SF accounting for the RM fluctuations induced by distance differences, which are independent of the separation $R$ between a pair of lines of sight \citep{LP16,Xup20}. The scaling of turbulence can still be revealed as long as the RM fluctuations induced by interstellar turbulence, which increase with $R$, can dominate over those induced by distance differences over a range of length scales.

There are other methods for measuring the anisotropy of turbulent magnetic fields and velocities. 
For instance, the anisotropy of turbulent magnetic fields can be extracted from synchrotron intensity and polarization data 
\citep{LP12,LP16}. 
{The sturcture function and quadrupole ratio modulus can be used to quantify the anisotropy of the synchrotron polarization intensity
\citep{LeC19,Waz20}.}
In addition, in spatially-coincident synchrotron emission and Faraday rotation regions, 
the statistical properties of the plane-of-sky and LOS components of magnetic fields can be separately obtained at different
wavelengths
\citep{LP16,ZhJ18}. 
The application of our method to studying the anisotropy of RM fluctuations in interstellar synchrotron-emitting media deserves 
further investigation. 
We note that although RMs are related to the LOS magnetic fields, the anisotropy obtained from RM fluctuations reveals the properties of the 
plane-of-sky mean magnetic field. 
\citet{Kan16} 
extended the VCA to study the anisotropy of turbulent velocities 
from velocity channel maps, which can be applied to supersonic MHD turbulence. 
The synergy of the above different methods,
as well as {other techniques for measuring magnetization, e.g., 
the VGT
\citep{Lakv18},
the synchrotron intensity and polarization gradients
\citep{Lazsyn18,Carm20},}
is necessary for achieving a comprehensive picture of the interstellar magnetic fields 
and turbulence.

\section{Summary}
\label{sec:5}
We have studied the anisotropy of SFs of RMs and VCs in  
the global reference frame of the mean magnetic field by using a set of 3D simulations of supersonic and sub-Alfv\'{e}nic MHD turbulence. 
Due to the overwhelmingly large contributions from isotropic density fluctuations in supersonic MHD turbulence, the turbulence anisotropy measured with RMs and VCs is significantly suppressed.This result is consistent with earlier findings in observations and simulations
(e.g., \citealt{XZ16,Hu20}). 

By selectively sampling the turbulence volume using the lines of sight with relatively low column densities, we find that the SFs of RMs and VCs normalized by column densities are both anisotropic, with the 
SFs measured in the direction perpendicular to the mean magnetic field larger than those measured parallel to the mean magnetic field. 
The anisotropy decreases with increasing $\rm M_A$ as $\rm M_A^{-4/3}$.
This is consistent with our theoretical expectation for anisotropic turbulent magnetic fields and velocities in sub-Alfv\'{e}nic MHD turbulence. The numerical result also shows that the anisotropy of 2D SFs has the same dependence on $\rm M_A$ as the 3D SF of turbulent velocities studied in \citet{Hua20}. We see that by selectively sampling the relatively diffuse regions in supersonic turbulence in cold interstellar phases, the anisotropic fluctuations of RMs and VCs can be used to determine the orientation of the POS magnetic field and its strength if the density is known.

By applying the radiative transfer calculations to our supersonic MHD simulations, we produce synthetic observations of $^{12}$CO and C$^{18}$O emission. In the presence of small-scale high-density structures in supersonic turbulence, the SF of normalized VCs steepens toward smaller scales due to the optical depth effect. As expected, the anisotropy obtained with the low-density tracer $^{12}$CO has the $\rm M_A$ dependence consistent with $\rm M_A^{-4/3}$. It can be used to measure both the direction and strength of the POS magnetic field.  The anisotropy measured with the higher-density tracer C$^{18}$O is smaller and has a weaker dependence on $\rm M_A$, which can still indicate the direction of the POS magnetic field. 

\acknowledgments
S.X. acknowledges the support for 
this work provided by NASA through the NASA Hubble Fellowship grant \# HST-HF2-51473.001-A awarded by the Space Telescope Science Institute, which is operated by the Association of Universities for Research in Astronomy, Incorporated, under NASA contract NAS5-26555. Y.H. acknowledges the support of the NASA TCAN 144AAG1967. We acknowledge the allocation of computer time by the Center for High Throughput Computing (CHTC) at the University of Wisconsin.
\software{MATLAB \citep{MATLAB:2018}, ZEUS-MP/3D code \citep{2006ApJS..165..188H}, Paraview \citep{Ahrens2005ParaViewAE}}

\bibliographystyle{aasjournal}
\bibliography{xu}

\end{document}